%% file: Bss-renorm.tex
\documentclass[11pt,a4paper]{article}
\pdfoutput=1   
\usepackage[utf8]{inputenc}
\usepackage{tabularx}
\usepackage{alpha} 
\hypersetup{%
   pdftitle    = {Precision renormalisation and improvement of Nf=3 lattice QCD with Wilson fermions},
   pdfauthor   = {P.Fritzsch, J.Heitger, S.Kuberski, R.Sommer, H.Simma},
   pdfkeywords = {PCAC, Lattice QCD, Nonperturbative Effects, O(a) improvement, HQET}
}%
\usepackage{lmodern}
\usepackage[outercaption]{sidecap}
\usepackage{macros}
\usepackage{xspace}
\usepackage{defs}
\usepackage{subcaption}
\usepackage{cancel}
\begin{document}
\input{title.tex}

\tableofcontents

\input{s_intro.tex}

\input{s_renorm_light.tex}

\input{s_renorm_heavy.tex}

\input{conclusion.tex}

\begin{acknowledgement}%
We are indebted to our colleagues in the ALPHA collaboration for many years of
fruitful collaboration, especially to Gregorio Herdo\'{i}za and Stefan Schaefer for a critical
reading of parts of the manuscript. P.F.\ was supported by the STFC Consolidated Grant
ST/P000479/1. S.K.\ and J.H.\ were supported by the Deutsche
Forschungsgemeinschaft (DFG) through the former Research Training Group GRK
2149. This project has received funding from the European Union’s Horizon
Europe research and innovation programme under the Marie Sk\l{}odowska-Curie
grant agreement No.\ 101106243. We gratefully acknowledge the Gauss Centre for
Supercomputing e.V. (\url{www.gauss-centre.eu}) for funding this project by
providing computing time on the GCS Supercomputer SuperMUC at the Leibniz
Supercomputing Centre (\url{www.lrz.de}). We furthermore acknowledge the
computer resources provided by the CIT of the University of Münster (PALMA II
HPC cluster) and thank its staff for support.
\\
Statistical uncertainties are determined and propagated using the
$\Gamma$-method \cite{Wolff:2003sm} in the implementations of the
\texttt{ADerrors.jl}~\cite{{Ramos:2018vgu,Ramos:2020scv}} and
\texttt{pyerrors}~\cite{Joswig:2022qfe} packages. 
\end{acknowledgement}
    
\appendix
  
\input{a_tabresfit.tex}

\input{a_compZA.tex}

\input{vectorcurrent.tex}
\input{variance.tex}

\input{app_sim.tex}

\small
\addcontentsline{toc}{section}{References}
\input{Bss-renorm.bbl}

\end{document}

%% file: title.tex
\preprintno{%
CERN-TH-2026-132\\
DESY-26-071\\
MS-TP-26-16\\
TCDMATH-26-01\\
}

\title{%
Precision renormalisation and improvement of\\
$\nf=3$ lattice QCD with Wilson fermions}

\collaboration{\includegraphics[width=2.8cm]{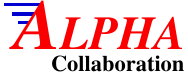}}

\author[tcd]{Patrick~Fritzsch}
\author[ms]{Jochen~Heitger}
\author[cern]{Simon~Kuberski}
\author[desy]{Hubert~Simma}
\author[desy,hu]{Rainer~Sommer}

\address[tcd]{School of Mathematics \& Hamilton Mathematics Institute, Trinity College Dublin, Dublin 2, Ireland}
\address[ms]{Institut für Theoretische Physik, Universität Münster, Wilhelm-Klemm-Str.\ 9, 48149 Münster, Germany}
\address[cern]{Theoretical Physics Department, CERN, 1211 Geneva 23, Switzerland}
\address[desy]{John von Neumann-Institut für Computing NIC, Deutsches Elektronen-Synchrotron DESY,\\Platanenallee 6, 15738 Zeuthen, Germany}
\address[hu]{Institut für Physik, Humboldt-Universität zu Berlin, Newtonstr.\ 15, 12489 Berlin, Germany}

\begin{abstract}
We renormalise (and improve) the flavour non-singlet axial current,
pseudo-scalar density, vector current and tensor current, as well as quark
masses, in $\rmO(a)$ improved lattice QCD with three massless flavours and
lattice spacings down to $0.01\,\fm$.
To this end, we tune a number of lattices with 
Schrödinger functional boundary conditions and resolutions $8\leq L/a\leq64$
to lines of constant physics with massless quarks and fixed gradient flow
coupling $\gbar_\mathrm{GF}^2(L_i),\; i=0,1,2$, corresponding to
$L_0\approx0.25\,\fm$, $L_1=2L_0$ and $L_2=4L_0$. We further renormalise and
improve the quark mass of additional heavy quarks for use in the B-physics
programme of the collaboration~\cite{Conigli:2023trw}. Our somewhat technical
results enable first-principles strategies for solving multi-scale problems
involving, e.g., the b-quark mass~\cite{Conigli:2023rod} or a large
temperature~\cite{Bresciani:2025vxw}. Comparing also to other determinations of
the axial current renormalisation constant $\za$, we have a precise
confirmation of how renormalisation and the restoration of chiral symmetry work
out with Wilson fermions at small $a$. In particular, the accurate restoration
of chiral symmetry and the exact flavour symmetry lead to practically
negligible uncertainties in observables determined from Ward identities: four
to five significant digits are achieved for $\za,\zv$. We provide an
explanation for the strong suppression of their statistical variances.
\newline\noindent
\end{abstract}


\maketitle

%% file: s_intro.tex
\section{Introduction}\label{s:intro}

Across the energy scales, QCD displays very different phenomena from light
hadron physics at low energy to the perturbative region in terms of weakly
interacting quarks and gluons at asymptotically high energy. Exploiting the
full predictive power of the fundamental formulation of the theory, lattice
QCD, requires connecting these regimes. Since available numerical resources do
not allow largely different energy scales to be treated in a common lattice
``simulation'', the ALPHA collaboration has developed the step-scaling
technique. It connects energy scales iteratively. In addition to the
renormalisation problems motivating the original development
\cite{Luscher:1991wu,Jansen:1995ck}, we would like to mention mastering aspects
of QCD at high temperature~\cite{Bresciani:2025vxw} as well as our approach to
treating the large b-quark mass scale in lattice
simulations~\cite{Heitger:2003nj,Guazzini:2007ja,Conigli:2023trw}.  We refer to
reviews \cite{Luscher:1998pe,Sommer:2015hea,DellaMorte:2015yda} as well as to
the most advanced application~\cite{Brida:2025gii} for an account of the
historical development and more references.

The basic idea \cite{Luscher:1991wu} is to introduce a finite, \emph{small}
volume with linear size $L$ where $L/a=\rmO(10)$ allows for small lattice
spacings, $a$, and therefore large energy scales. The finite-volume physics is
related iteratively, $L_0\to L_1=2L_0\to L_2=2L_1=2^2L_0$ (and, depending on the
application, also more steps), to the large-volume physics of interest. We note
that in this process the larger but finite volumes with $L\approx 1\,\fm$ are
particularly challenging, as the guidance from perturbation theory can be
grossly misleading~\cite{DellaMorte:2008xb}.

Here we carry out precise simulations with $L=L_0,L_1,L_2$, as specified above,
with $L_0\approx \frac{1}{4}\,\fm$ and three massless flavours of $\rmO(a)$ improved
Wilson quarks. We compute the improvement and renormalisation coefficients,
which are needed to renormalise the \textit{charm} and \textit{bottom} quark
masses as a prerequisite for the strategy described in
Ref.~\cite{Conigli:2023trw} and already used in our preliminary
reports~\cite{Conigli:2023rod,Kuberski:2026bpx}. We further compute
renormalisation factors for the flavour non-singlet axial-vector current, the
pseudo-scalar density, as well as the vector and tensor currents. In the case of
the axial and vector currents, we also provide results for the coefficients of
the associated $\rmO(a)$ improvement and mass-improvement terms, respectively. Our
simulations of the finite-volume theory with Schr\"odinger functional boundary
conditions~\cite{Luscher:1992an,Sint:1993un} set a new standard in terms of the
resolutions achieved. We go up to $L/a=64$, working with lattice spacings
between $0.007\,\fm$ and $0.08\,\fm$. 

In this paper we describe our simulations, including the non-trivial setup of
the proper lines of constant physics and the determination of the above-mentioned 
coefficients. In addition, we present step-scaling functions for the
renormalised coupling and quark mass, improving further on the results of
\cite{DallaBrida:2016kgh,Campos:2018ahf} in terms of precision and resolution.
The small lattice spacings allow a check that the continuum limit is achieved
smoothly and the extrapolations of \cite{DallaBrida:2016kgh,Campos:2018ahf}
were safe. Similarly, we check and advance the improvement and
renormalisation of the axial current. We demonstrate rather precisely how
different renormalisation conditions yield compatible results once their
intrinsic discretisation errors are accounted for. All quantities determined
by Ward identities (WIs) are very precise when the lattice spacing is small.
We explain in appendix~\ref{app:variance} that this is due to the fact that,
close to the continuum, the statistical variance of quantities determined from
WIs is strongly dominated by contact terms and these are typically suppressed
by $g_0^2$- and $a/L$-factors. 

This paper is organised as follows. In section~\ref{s:massless} we present the
non-perturbative renormalisation and improvement in the massless theory. After
defining the line of constant physics and the tuning procedure, we discuss the
renormalisation of the axial current and pseudo-scalar density, and introduce
the step-scaling framework. We also revisit the improvement of the axial
current in this setup. In section~\ref{s:massive} we extend the analysis to
massive quenched flavours, including the renormalisation of quark masses and
the formulation of a partially massive scheme. Our main conclusions are
summarised in section~\ref{s:conclusion}. Additional results, tables and fit
formulae are collected in appendix~\ref{a:tables}.
Appendix~\ref{sec:ambiguitiesZA} compares different definitions of $\za$. The
renormalisation of the vector and tensor currents is discussed in
appendix~\ref{a:zvt}. Appendix~\ref{app:variance} contains an analysis of the
statistical variances of PCAC quark masses. Details on the gauge ensembles and
simulation algorithms, as well as the scaling of autocorrelation times, are
provided in appendix~\ref{a:ensembles}.

%% file: s_renorm_light.tex
\section{Renormalisation of the unitary theory}\label{s:massless}

\subsection{Line of constant physics}

For our step-scaling applications we need lattices with a finite and fixed size
$L=L_0,L_1,L_2$. It is convenient to choose vanishing quark masses and
Schr\"odinger functional (SF) boundary
conditions~\cite{Luscher:1992an,Sint:1993un}. Continuum extrapolations are
enabled by lattices with a wide range of resolutions $a/L$.  For each chosen
$a/L$, we then tune the bare parameters, $g^2_0,\,m_0$ such that the quark mass
vanishes (for a proper definition of the quark mass at finite lattice spacing)
and a finite-volume renormalised coupling $\bar{g}_\mathrm{GF}^2(L)$ has a
fixed value. The set of bare parameters implementing these conditions is
denoted a ``line of constant physics'' (LCP).

We follow ref.~\cite{DallaBrida:2016kgh} concerning details: we set fermionic
phases in the spatial boundary conditions  of each quark to $\theta_k = \theta
= 0.5$ ($k=1,2,3$) and work with vanishing boundary gauge fields.  $\rmO(a)$
improvement at the boundaries is achieved by choice~B of
Ref.~\cite{Aoki:1998qd} together with the one-loop formulae for the boundary
improvement coefficients $\ct$ and $\tilde{c}_t$ from
refs.~\cite{Sint:1995ch,Takeda:2003he,DallaBrida:2016kgh}. We define our first
LCP by a renormalised coupling $\bar{g}_\mathrm{GF}^2(L_0) = 3.949$ at
vanishing quark mass, $m=0$. We label it LCP-0 and note that $L_0\approx
\frac{1}{4}\,\fm$ can be deduced from \cite{DallaBrida:2016kgh,Bruno:2017gxd};
the other LCPs (cf. tables \ref{tab:ens_2L0}, \ref{tab:ens_HQET1},
\ref{tab:ens_H2s}, \ref{tab:ens_HQET2}, \ref{tab:ensL02T}) will be discussed as
we go along. We refer to appendix~\ref{a:ensembles} for further details on the
gauge ensembles and simulation setup.  For convenience and later use we now
give the precise definition of the coupling and the quark mass.

\subsubsection{Coupling}
The finite-volume gradient flow coupling is defined as in
\cite{DallaBrida:2016kgh} with flow time $t=c^2L^2/8$ and the choice $c=0.3$,
\begin{align}\label{eq:g2GF}
        \bar{g}_\mathrm{GF}^2(L) &= \left. \mathcal{N}^{-1}(a/L) {\,t^2\, \langle E_\mathrm{mag}(t,L/2) \rangle_{Q=0}}\right|_{\sqrt{8t}=0.3L}\,,
\end{align}
where  
\begin{align}
  E_\mathrm{mag}(t,x_0) 
  = \frac{a^3}{L^3} \sum_\mathbf{x}\left[\frac{1}{4} G_{ij}^a(t,x)G_{ij}^a(t,x) \right]^\mathrm{LW}
  = \frac{a^3}{L^3} \sum_\mathbf{x}\left[\frac{5}{3} E_\mathrm{mag}^\mathcal{P} (t, x) -\frac{1}{12}E_\mathrm{mag}^\mathcal{R}(t, x)\right] 
\end{align}
is the magnetic part ($i,j=1,2,3$) of the action density in the Lüscher--Weisz
gauge action,
\begin{align}
E_\mathrm{mag}^\mathcal{X}(t, x)=-\frac{1}{2a^4}{\sum}_{i,j}\left[ \tr\big( \mathcal{X}_{ij}(t,x)+ \mathcal{X}_{ij}(t,x)^\dagger\big)-2N  \right] \;. 
\end{align}
It is an $\rmO(a^2)$ improved discretisation of  
$E_\mathrm{mag}^\mathrm{cont}(t,x_0) = \frac{1}{4L^3} {\displaystyle\int}G_{ij}^a(t,x)G_{ij}^a(t,x)\,\mathrm{d^3}x$ 
made of (spatial) plaquettes $\mathcal{P}$ and $2\times 1$ rectangles $\mathcal{R}$,
evolved to positive flow times using the Zeuthen flow~\cite{Ramos:2015baa}.
The normalisation $\mathcal{N}(a/L)$~\cite{Fritzsch:2013je,Ramos:2015baa} ensures
$\bar{g}_\mathrm{GF}^2=g_0^2+\rmO(g_0^4)$.  The coupling thus has no
discretisation errors at leading order in perturbation theory.  The relevant
values of $\mathcal{N}$ for our definition of the coupling at $c=0.3$ are
quoted in table~\ref{tab:norm_g2GF}.

The definition of our finite-volume observables includes a projection onto the
sector of vanishing topological charge~\cite{Fritzsch:2013yxa} in order to
circumvent problems with topology freezing in the simulations. We denote the
projected path integral expectation values by $\langle\bullet\rangle_{Q=0} $
and define them in appendix~\ref{sec:Q}.

\subsubsection{Current (PCAC) quark masses}

With Wilson fermions, quark masses are best defined through the PCAC
relation%
\footnote{$\partial_\mu,\partial_\mu^*$ denote the usual forward and backward
          derivatives, and $\tilde{\partial}_\mu$ their average.
}
\begin{align}
	\langle [\tilde{\partial}_0 A^{ij}_0(x)+ac_\mathrm{A} \partial_0^*\partial_0 P^{ij}(x)]  \, \mathcal{O}^{ji} \rangle_{Q=0} &= 2 m_{ij}(x_0)\langle P^{ij}(x) \mathcal{O}^{ji}\rangle_{Q=0} \, ,
	\\
	m_{ij} &\equiv \frac12 (m_i+m_j) + \rmO(a^2)\,,
\end{align}
based on the axial current and pseudo-scalar density, given for distinct
flavours $i\neq j$ by
\begin{align}
	A_\mu^{ij}(x) = \psibar_i(x) \gamma_\mu \gamma_5 \psi_j(x)\,,\qquad P^{ij}(x) = \psibar_i(x) \gamma_5 \psi_j(x) \,,
\end{align}
respectively. The axial current
is $\mathrm{O}(a)$ improved in the chiral limit with the
coefficient $c_\mathrm{A}(g_0^2)$, which has been non-perturbatively determined
for our action in ref.~\cite{Bulava:2015bxa}. In section~\ref{sec:ca}, we also
present a determination of $\ca$ from our simulations that confirms the
interpolation of ref.~\cite{Bulava:2015bxa} between low energy and the one-loop
perturbative expression.

We use 3-momentum $\mathbf{p}=0$ and pseudo-scalar fields $\mathcal{O}^{ji},
\mathcal{O}'^{ji}$ defined at the boundaries $x_0=0,T$ of the SF, respectively.
In terms of the dimensionless boundary quark fields \cite{Luscher:1996sc}, they read%
\begin{eqnarray}\label{e:boundopstat}
    {\mathcal{O}}^{ji} ={a^6\over L^3} \sum_{\vecu,\vecv}\zetabar_j ({\vecu}) \gamma_5 \zeta_i({\vecv}) \,,&& 
    {\mathcal{O}}'^{ji}={a^6\over L^3} \sum_{\vecu,\vecv}\zetabar'_j({\vecu})\gamma_5\zeta'_i ({\vecv}) \,. 
\end{eqnarray}
Flavour indices $1\leq i,j \leq3$  label our three mass-degenerate sea quarks.
The exact flavour symmetry implies
\begin{eqnarray}
	m_{12}(x_0)=m_{13}(x_0)=m_{23}(x_0) \equiv m(x_0) \,.
\end{eqnarray}
We will later use additional quenched massive valence quarks with indices $i,j
\geq4$.  Since the PCAC relation is a chiral Ward identity, the
$x_0$-dependence of $m_{ij}(x_0)$ is only a discretisation effect.  Such
effects are enhanced close to the SF boundaries. We avoid those but still
improve the statistical precision by the definition
\begin{align}
        m_{ij} &\equiv \dfrac{1}{|I_{T}|} \sum_{x_0/a\,\in I_T} m_{ij}(x_0) \,, &
        I_T &= \Big\{ \frac{3T}{8a},\ldots,\frac{T}{2a},\ldots,\frac{5T}{8a} \Big\} \,,
        \label{e:x0average}
\end{align}
where we round up to the nearest integer, if necessary. In physical units,
$I_{T}$ is  constant up to rounding effects when $a$ is varied at fixed $T$.
%

\subsubsection{Tuning results}

The tuning towards vanishing quark mass has been significantly simplified by
using the information on the critical line provided by
ref.~\cite{DallaBrida:2016kgh} as a first approximation. With few iterations
we achieved 
\begin{equation}
	|Lm| < 0.001\,,
\end{equation}
on all except one ensemble, as documented in table~\ref{tab:ens_L0}, and the
coupling is tuned to $3.949$ within statistical uncertainties of at most
$0.01$. This corresponds to a precision on the renormalised quark mass of
0.4~MeV and on $L_0$ of 0.5\%.
\begin{table}[t]
	\setlength{\abovecaptionskip}{10pt}
	\setlength{\belowcaptionskip}{0pt}
	\centering\small
	\renewcommand{\arraystretch}{1.25}
    \caption{Ensembles tuned to {\bf LCP-0} ($\gbar^2_\mathrm{GF}(L)=3.949$,
             i.e., $L=L_0$, and vanishing quark mass).  All ensembles feature
             $T=L$. $N_{\rm cfg}$ is the total number of configurations,
             separated by $\tau_{\rm ms}$ molecular dynamics (MD) units, on
             which the coupling and Schrödinger functional correlation functions
             are evaluated to compute the sea current quark mass $am=am_{12}$
             and renormalisation factors $\ZA$ and $\ZP$. The MD
             trajectory/integration length is $\tau=2$\,MDU.  Note that $N_{\rm
             cfg}=N_0$, the number of configurations with vanishing topological
             charge $Q$, because on these very fine lattices in small volume the
             charge is frozen to zero and thus the $Q=0$ projection
             (appendix~\ref{sec:Q}) has no effect here.  
            }
	\input{./tables/enstab_L0.tex}
	\label{tab:ens_L0}
\end{table}

We are also interested in non-zero---indeed, large---renormalised quark masses.
Starting from the above PCAC masses we just need to renormalise the
(non-singlet) axial current and the pseudo-scalar density.  We turn to their
renormalisation constants, $\za$ and $\zp$. 

\subsection{Axial current renormalisation}

In refs.~\cite{Bulava:2016ktf,DallaBrida:2018tpn}, $\za$ has already been
computed with $N_\mathrm{f}=3$ flavours using the very same action as in our
project. In particular \cite{Bulava:2016ktf},
following~\cite{Luscher:1996jn,DellaMorte:2005xgj}, determined $Z_\mathrm{A}$
enforcing an integrated chiral Ward identity with SF boundary conditions.
Their direct non-perturbative  results are for larger lattice spacings than the
ones we employ for $L=L_0,L_1$. For our smaller lattice spacings a fit function
interpolating the non-perturbative result and the perturbative asymptotic
behaviour is available, but its exact precision is not easily determined. We
thus perform a direct determination using the ensembles described above.
Similar to~\cite{Luscher:1996jn} we define
\begin{align}\label{e:ZA_WI}
    \za                 &= \lim_{m\to 0} \za^\mathrm{eff}(m)\,, &  
    \za^\mathrm{eff}(m) &= \left[F_1\Big/ \Big( F_\mathrm{AA}^\mathrm{I}(x_0, y_0) - 2m \cdot \tilde{F}_\mathrm{PA}^\mathrm{I}(x_0, y_0) \Big) \right]^{\frac{1}{2}} \,,
\end{align}
with sea current quark mass $m=m_{12}$ and improved correlation functions
\begin{align}
    F_\mathrm{AA}^\mathrm{I}(x_0, y_0) 
        &= F_\mathrm{AA}(x_0, y_0) 
        + a c_\mathrm{A}\left[\tilde{\partial}_{x_0} F_\mathrm{PA}(x_0, y_0) + \tilde{\partial}_{y_0}F_\mathrm{AP}(x_0, y_0)\right] \notag\\
        &\hspace*{2.45cm}+ a^2 c_\mathrm{A}^2 \tilde{\partial}_{x_0}\tilde{\partial}_{y_0}F_\mathrm{PP}(x_0, y_0) \,,\\
    \tilde{F}_\mathrm{PA}^\mathrm{I}(x_0, y_0) &= a \sum_{x'_0=y_0}^{x_0}w(x'_0)\left[F_\mathrm{PA}(x'_0, y_0) + a c_\mathrm{A} \partial_{y_0}F_\mathrm{PP}(x'_0, y_0)\right] \,.
\end{align}
The central difference operator $\tilde{\partial}$ and the implementation of
the trapezoidal rule via
\begin{align}
    w(x'_0) = \begin{cases}     
                1/2 & \text{if } x'_0 = y_0 \text{ or } x'_0 = x_0\\
                1   & \text{if } y_0 < x'_0 < x_0
              \end{cases} 
\end{align}
are used. In eq.~\eqref{e:ZA_WI}, the SF four-point functions
$F_\mathrm{XY}$ are defined by
\begin{eqnarray}\label{eq:FXY}
F_{\rm XY}(x_0,y_0)=-{{\rmi a^6}\over{6}}\sum_{\bf x,y} \sum_{i,j=1}^2 \epsilon^{abc} \langle 
{\mathcal{O}}'^{\,3i} X^a(x)\,Y^b(y) (\frac{1}{2}\tau^c)_{ij} {\mathcal{O}}^{3j} \rangle \,,
\end{eqnarray}
with the totally antisymmetric tensor $\epsilon^{abc}$ (with $\epsilon^{123}=1$) 
and the boundary fields $\mathcal{O},\mathcal{O}'$ from
eq.~(\ref{e:boundopstat}), while the boundary-to-boundary correlator $F_1$
is
\begin{align}\label{e:f1}
    F_1\equiv-{{1}\over{8}}\,\sum_{i=1}^2\langle{\mathcal{O}}'^{3i} {\mathcal{O}}^{i3}\rangle \,. 
\end{align}
As a slight variation of~\cite{DellaMorte:2005xgj}, we have taken boundary
fields involving the strange quark (flavour 3) but the Ward identity following
from an SU(2) axial rotation in the $i,j\in\{1,2\}$ space. The Wick
contractions then contain only connected diagrams, and one obtains directly
what is labelled $Z_\mathrm{A}^\mathrm{con}$ in ref.~\cite{Bulava:2016ktf}.
In section~4.2 of~\cite{DellaMorte:2008xb}, the absence of disconnected
diagram contributions was achieved by using a static quark instead of the
strange quark.%
\footnote{While in \cite{Bulava:2016ktf} wave functions have been used in the
        definition of the boundary sources to reduce excited state
        contributions, we here work for smaller $T=L$, closer to the
        perturbative region, where the analysis of \cite{DellaMorte:2008xb}
        shows that  standard SF boundary sources are of advantage.
}

\begin{figure}
    \centering
    \includegraphics[width=15cm]{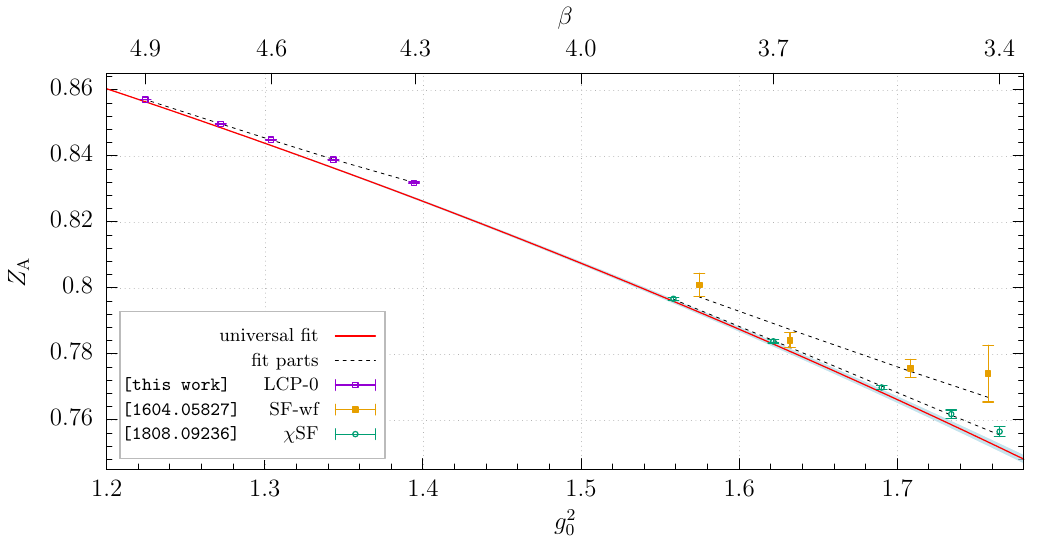}
    \caption{$\ZA$ determined along the LCP-0  compared to the determination
             from the chirally rotated SF~\cite{DallaBrida:2018tpn} and from the
             standard SF.  The latter is the version $\ZA^\mathrm{con}$ of
             \cite{Bulava:2016ktf}, which coincides with our definition apart
             from  the use of a non-trivial smearing function employed for the
             boundary sources, which enhances the ground state contribution. 
            }	
    \label{fig:ZA_comb}
\end{figure}

It was noted in ref.~\cite{Bulava:2016ktf} that the dependence of
$\za^\mathrm{eff}(m)$ on the quark mass is very mild. Our sea quarks are
massless to a very good approximation. Thus we do not need to carry out the
limit $m\to 0$ explicitly.
To maximise the distance between the insertion points, we choose
$x_0=\frac{2}{3}T$ and $y_0 = \frac{1}{3}T$ (possibly rounding $x_0/a,y_0/a$ to
the next integer). On our LCP-0 lattices, with $T=L$, the insertions are rather
close to the boundaries, namely, $y_0=\frac{1}{3}L$ is as small as four lattice
spacings. In appendix~\ref{sec:ambiguitiesZA}, we show that the difference to $T=2L$ as
used in the past is a {\em small} $(a/L)^2$-effect and---with our lattice
spacings---the choice $T=L$ is sufficient. We summarise the results for our
determination of $Z_\mathrm{A}$ in table~\ref{tab:ens_L0}. $\ZA$ has much
smaller statistical uncertainties than previous determinations. The small
uncertainties are a property of quantities determined by WIs as explained in
appendix~\ref{app:variance}. A comparison is shown in
figure~\ref{fig:ZA_comb}.
The uncertainties of our determination are not visible at this scale. The
curve labeled ``universal fit'' is determined in the following manner.
Different definitions of $\ZA$, with different LCPs, have to agree up to
discretisation errors of $\rmO(a^2)$. We thus attempt to describe all data
(with $\beta > 3.3$) by a universal fit function 
\begin{eqnarray}
	\ZA^\mathrm{univ}(g_0)  = 1+\ZA^{(1)} g_0^2 + c_2 g_0^4 + c_3 g_0^6\,,
	\quad \text{$\ZA^{(1)}=-0.090488$ (one-loop~\cite{Aoki:1998ar})}\,,
	\label{e:zauniv}
\end{eqnarray}
together with discretisation errors
\begin{eqnarray}
  \ZA(g_0) - \ZA^\mathrm{univ}(g_0)  =
  \begin{cases}
   	k_1 \cdot a^2/L^2  & \text{for LCP-0} \\
   	k_2 \cdot a^2/L^2  & \text{for $\ZA^\mathrm{con}$ of \cite{Bulava:2016ktf}} \\
   	k_3 \cdot g_0^4\,a^2/L^2  & \text{for $Z^l_\mathrm{A, sub}$ ($L_2$-LCP) of \cite{DallaBrida:2018tpn}}
  \end{cases}\,.
\end{eqnarray}
The extra factor $g_0^4$ for the cutoff effects in $Z^l_\mathrm{A, sub}$
accounts for the subtraction of one-loop discretisation errors
\cite{DallaBrida:2018tpn}. We obtain an excellent fit with just two fitted
powers of $g_0^2$ after quadratically adding a small systematic error of size
$(a/L)^3/2$ to all data points. The latter is a phenomenological account of
higher-order cutoff effects. The systematic error is rather significant for the
three largest $a/L$ of our very precise results (LCP-0) and still relevant for
the largest $a/L$ of~\cite{DallaBrida:2018tpn}. The universal fit parameters
are 
\begin{equation}
  (c_2,c_3) = (-0.00693815,  \;-0.0122132)\,,
\end{equation}
with their covariance matrix
\begin{equation}
  \mathrm{cov} = \begin{pmatrix}
   +0.5761  & -0.4860 \\
   -0.4860  & +0.4107		
	\end{pmatrix} \times 10^{-6}\,.
\end{equation}

In appendix~\ref{sec:ambiguitiesZA}, figure~\ref{fig:ZA_diff}, we show the
quality of the global fit in more detail.  Figure~\ref{fig:ZA_comb} and
figure~\ref{fig:ZA_diff} represent a remarkable demonstration how
renormalisation of the axial current works out with Wilson fermions and how
chiral symmetry is restored in the continuum limit.

For future use, eq.~\eqref{e:zauniv} is a perfectly legitimate choice for all
$g_0^2\leq 1.78$. However, if the bare couplings match one of the direct
determinations, those results are still to be preferred.  In particular the
numbers in table~\ref{tab:ens_L0} are more precise than the fit function, since
the latter does contain an uncertainty contribution from the added
$(a/L)^3$-term.

\subsection{Renormalisation of the pseudo-scalar density}\label{app:ZP}

The renormalised pseudo-scalar density is scale-dependent. Implementing a
mass-in\-de\-pen\-dent SF scheme, we define the renormalisation constant
$Z_\mathrm{P}(L)$ in the massless theory and in finite volume of size
$L^4$~\cite{Capitani:1998mq, Campos:2018ahf}. The linear extent $L$  plays the
role of the renormalisation scale, $\mu=1/L$. As for $Z_\mathrm{A}$, the
constant $Z_\mathrm{P}$ additionally depends on the bare coupling $g_0$, while
specifying $L$, e.g., by $L=L_0$, the resolution $a/L_0$ is fixed, cf.\
table~\ref{tab:ens_L0}.  In terms of $F_1$, eq.~(\ref{e:f1}), and the
boundary-to-bulk correlation function 
\begin{eqnarray}
    f_\mathrm{P}(x_0)=-{{1}\over{2}}\,a^3\sum_{\bf x}\langle P^{12}(x){\mathcal{O}}^{21}\rangle \,,
\end{eqnarray}
the renormalisation constant $Z_\mathrm{P}$ is defined as
\begin{align}\label{eq:ZPdef}
        Z_\mathrm{P} = c\cdot \left. {\sqrt{3F_1}}\Big/{\fP(T/2)}\right|_{m=0}\,,
\end{align}
where the normalisation constant $c$ is chosen such that $Z_\mathrm{P}=1$ at
tree-level.  Numerical values for $c(L/a)$ for our setup with $T=L$, $\theta =
0.5$ and vanishing boundary gauge fields are listed in table~\ref{tab:ZP_norm}.
We include our non-perturbative results for $Z_\mathrm{P}(L_0)$ in
table~\ref{tab:ens_L0}. They have per-mille level precision.

The renormalisation of the vector and tensor currents, which proceeds along
similar lines as for the axial current and pseudo-scalar density, respectively,
is discussed in appendix~\ref{a:zvt}.

\subsection{Step-scaling}

In order to implement the ALPHA collaboration strategy for B-physics
\cite{Bernardoni:2013xba,Conigli:2023trw,Conigli:2023rod}, we simulated
step-scaling lattices which implement 
\begin{equation}
	L=2L_0\approx 0.5\,\fm\,,\quad L=2L_1\approx 1\,\fm\,,
\end{equation}
with precise definitions
\begin{equation}
  \gbsq_\mathrm{GF}(L_0)\equiv u_0=3.949\,, 	\quad \gbsq_\mathrm{GF}(L_1)\equiv u_1=5.867\,.
\end{equation}
Here the chosen value for $\gbsq_\mathrm{GF}(L_1)$ is determined from
$\gbsq_\mathrm{GF}(L_0)$, exploiting the continuum step-scaling function,
\begin{eqnarray}
	\sigma_\mathrm{GF}(u) = \left.\gbsq_\mathrm{GF}(2L)\right|_{\gbsq_\mathrm{GF}(L)=u,m=0}\,,
\end{eqnarray} 
with numerical value $\sigma_\mathrm{GF}(3.949)=5.867(29)$
from~\cite{DallaBrida:2016kgh}.  We tuned to $\gbsq_\mathrm{GF}(L_1)$ with a
statistical precision of $\pm 0.014$ and central values matching to $\pm 0.01$,
cf.\ table~\ref{tab:ens_HQET1} in appendix~\ref{sec:lcps}. For the contact to
infinite-volume lattices of CLS, there is also a set of ensembles tuned to 
\begin{eqnarray}
	\gbsq_\mathrm{GF}(L_2)=11.27\,,
\end{eqnarray}
listed in table~\ref{tab:ens_HQET2}. All LCPs are visualised in the
$(g_0^2,L/a)$-plane in figure~\ref{fig:LCPs}.
\begin{table}[t!]
	\setlength{\abovecaptionskip}{10pt}
	\setlength{\belowcaptionskip}{0pt}
	\renewcommand{\arraystretch}{1.25}
	\centering\small
    \caption{Step-scaling ensembles {\bf LCP-0s} with bare gauge coupling and
            hopping parameters of LCP-0, table~\ref{tab:ens_L0}.  We
            list the measured gradient flow coupling $\gbar^2_{\rm GF}$ and
            current sea quark mass $am=am_{12}$.
            $N_{\rm cfg}$ is the number of configurations, separated by
            $\tau_{\rm ms}$ molecular dynamics (MD) units. The trajectory
            length is $\tau=2$ throughout.
 	                }\label{tab:ens_2L0}%
	\input{./tables/enstab_2L0.tex}
\end{table}
\begin{figure}[t!]
	\setlength{\abovecaptionskip}{10pt}
	\setlength{\belowcaptionskip}{0pt}
	\centering\small
	\renewcommand{\arraystretch}{1.25}
    \includegraphics[width=0.95\textwidth]{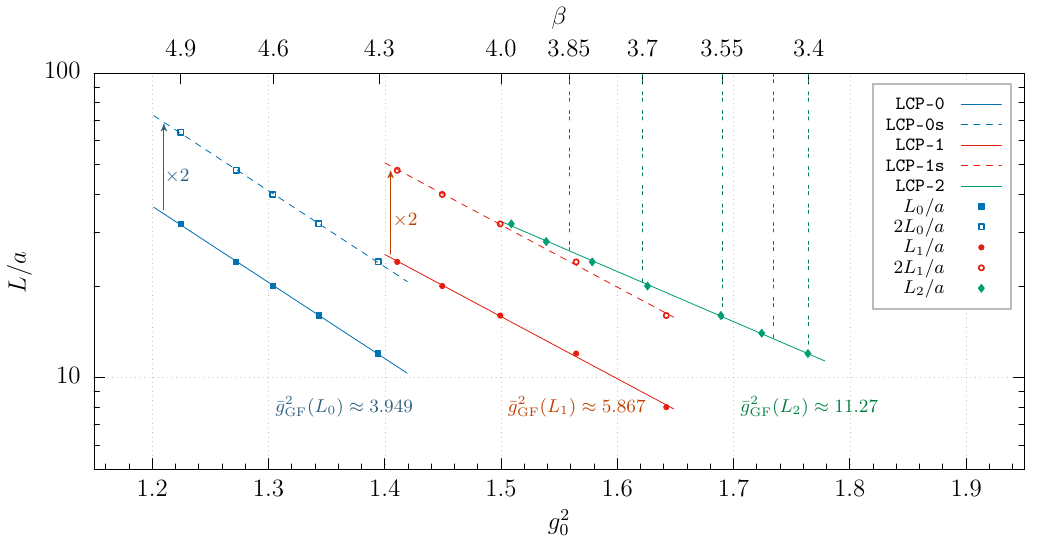}
    \caption{Relation between bare coupling $g_0^2$ and simulated lattice
            sizes $L/a$, imposed through the constant renormalised coupling
            condition to define LCPs at $\mu^{-1}=L\in\{L_0,L_1,L_2\}$, cf.
            tables~\ref{tab:ens_L0}, \ref{tab:ens_HQET1}
            and~\ref{tab:ens_HQET2}.  Vertical dashed lines indicate couplings
            simulated by the CLS consortium, and curves are added to guide the
            eye.
	        }
    \label{fig:LCPs}
\end{figure}

For the steps $L_i\to L_{i+1}, \;i=0,1$, we simulated the doubled lattices at
the bare parameters of the determined LCPs implementing fixed $L_i$, cf.
tables~\ref{tab:ens_2L0} and~\ref{tab:ens_H2s}. Since the (light, unitary)
lines of constant physics have been fixed in $L_0$ and $L_1$, the degenerate
light quark masses almost vanish. Their small non-zero values are due to
discretisation effects and misstunings in $L_0$ and $L_1$. 

\begin{figure}[t!]
    \centering
    \includegraphics[width=.5\textwidth]{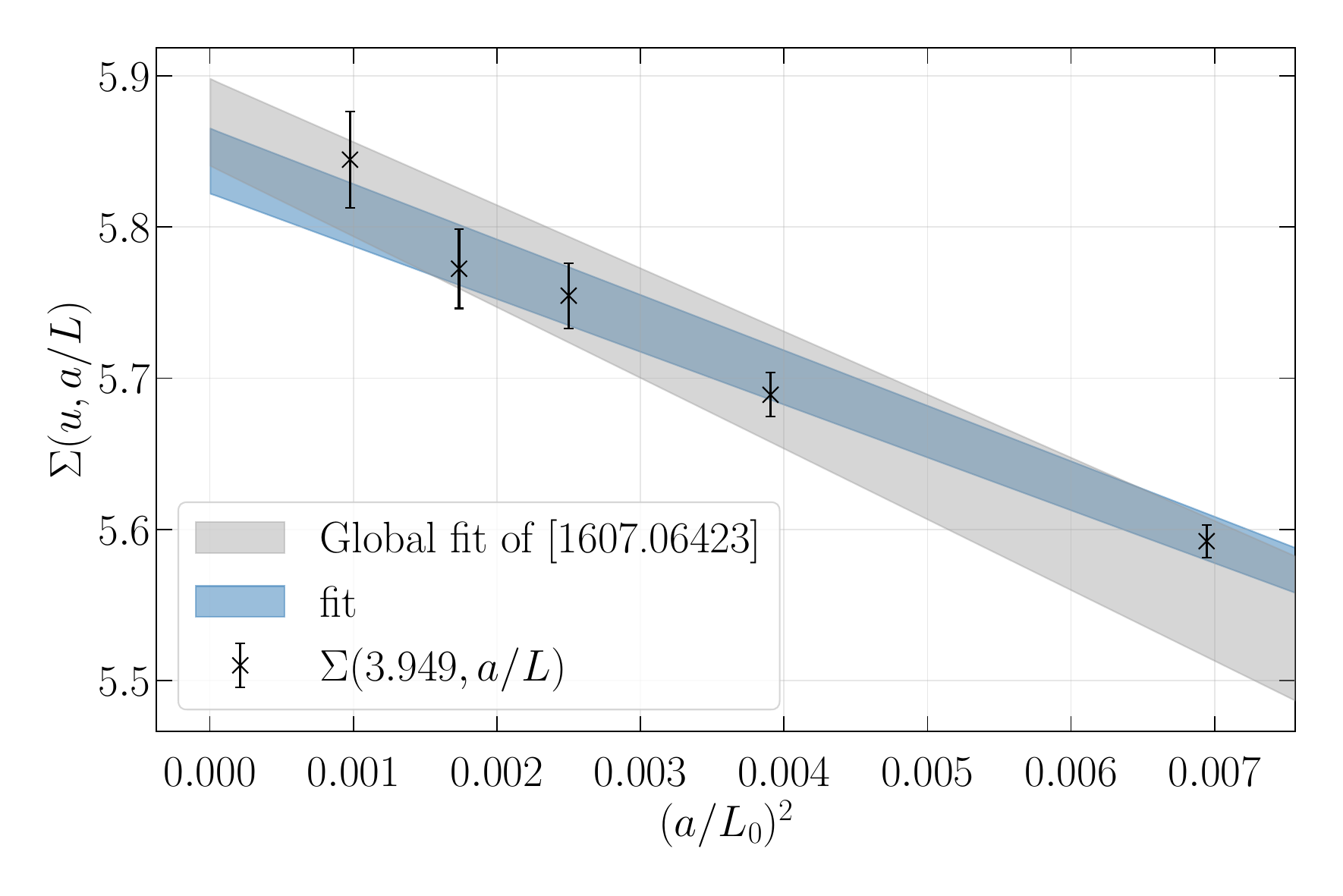}%
    \includegraphics[width=.5\textwidth]{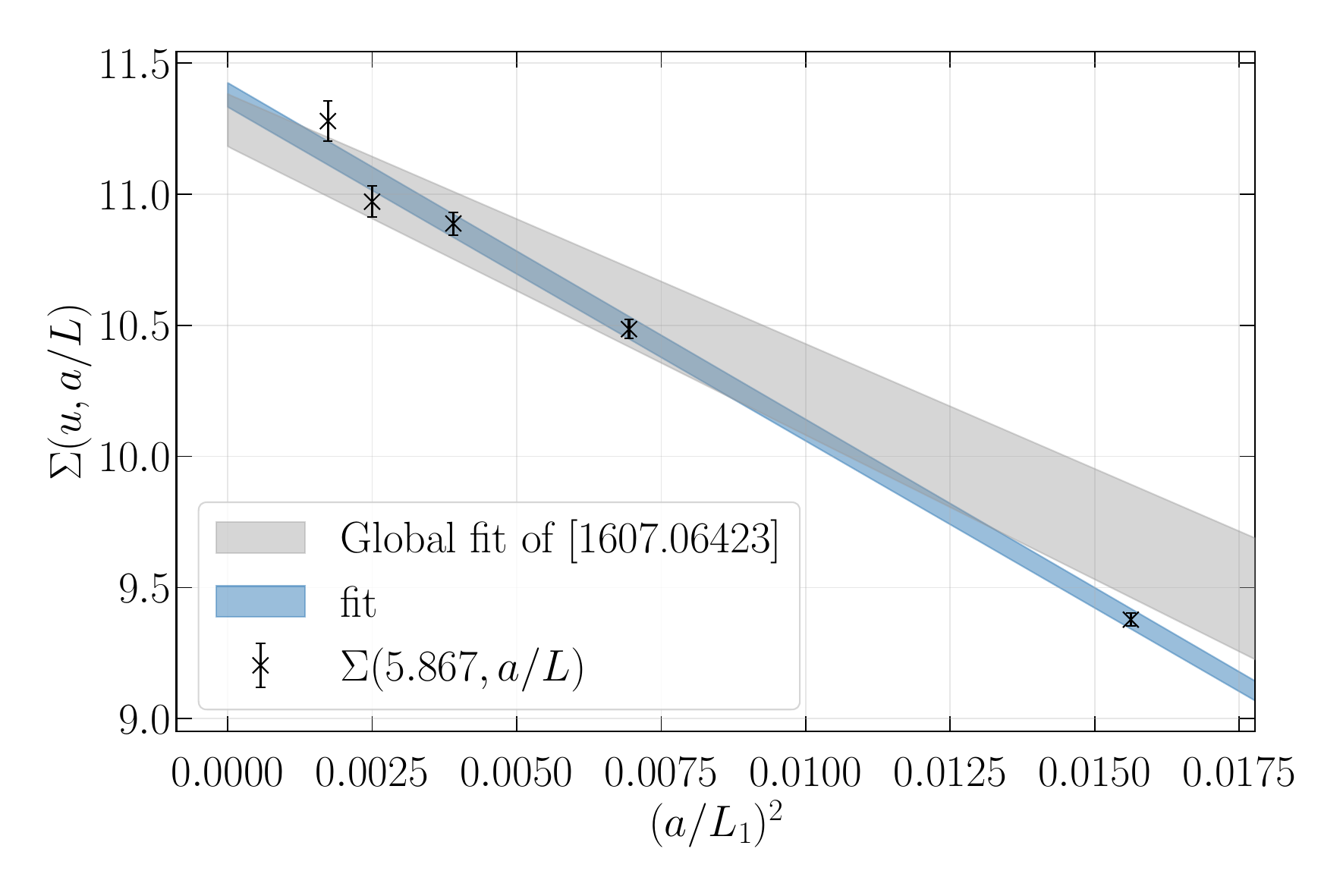}\\%
    \caption{Step-scaling functions for $L_0\to2L_0$ (left) and $L_1\to2L_1$
             (right). The gray error band shows the  global fit to these
             step-scaling functions performed in \cite{DallaBrida:2016kgh}. Note
             that this fit did not include the data points presented here. The
             lattice approximants $\Sigma(u,a/L)$ are shifted to the desired
             $u_0,u_1$ via the parameter-free one-loop formula
             $\Sigma(u_i,a/L)^{-1} = \Sigma(u,a/L)^{-1} + u_i^{-1} - (u)^{-1}$,
             where $u=\gbar^2_{\rm GF}(L)$ is the result of the simulations
             (tables~\ref{tab:ens_2L0} and \ref{tab:ens_H2s}). For the small
             shifts needed, the one-loop formula is very accurate.
            }
    \label{fig:gfstep}
\end{figure}

Fits of the form $\bar{g}_\mathrm{GF}^2(2L_i) = \sigma_\mathrm{GF}(u_i)  +
k(a/L_i)^2$ to the data of tables~\ref{tab:ens_2L0} and~\ref{tab:ens_H2s} yield
$\sigma_\mathrm{GF}(u_0) = 5.844(22)\,, \; \sigma_\mathrm{GF}(u_1)=
11.379(46)$. In ref.~\cite{DallaBrida:2016kgh} an analysis of a large set of
data with many values of $L$ was performed. Our continuum values agree with the
result of~\cite{DallaBrida:2016kgh} and, additionally, the data at finite $a/L$
are well described  by their global fit as shown by the grey bands in
figure~\ref{fig:gfstep}.  Since here we use significantly smaller $a/L$ and
achieve better statistical accuracy, this agreement is an excellent
confirmation of the analysis strategy of ref.~\cite{DallaBrida:2016kgh}.
Indeed, an analysis of the two data sets combined \cite{Brida:2025gii} has
contributed to a very precise determination of the strong coupling
$\alpha_{\overline{\mathrm{MS}}}(m_\mathrm{Z})$.
\begin{figure}[t!]
	\centering
	\includegraphics[width=.5\textwidth]{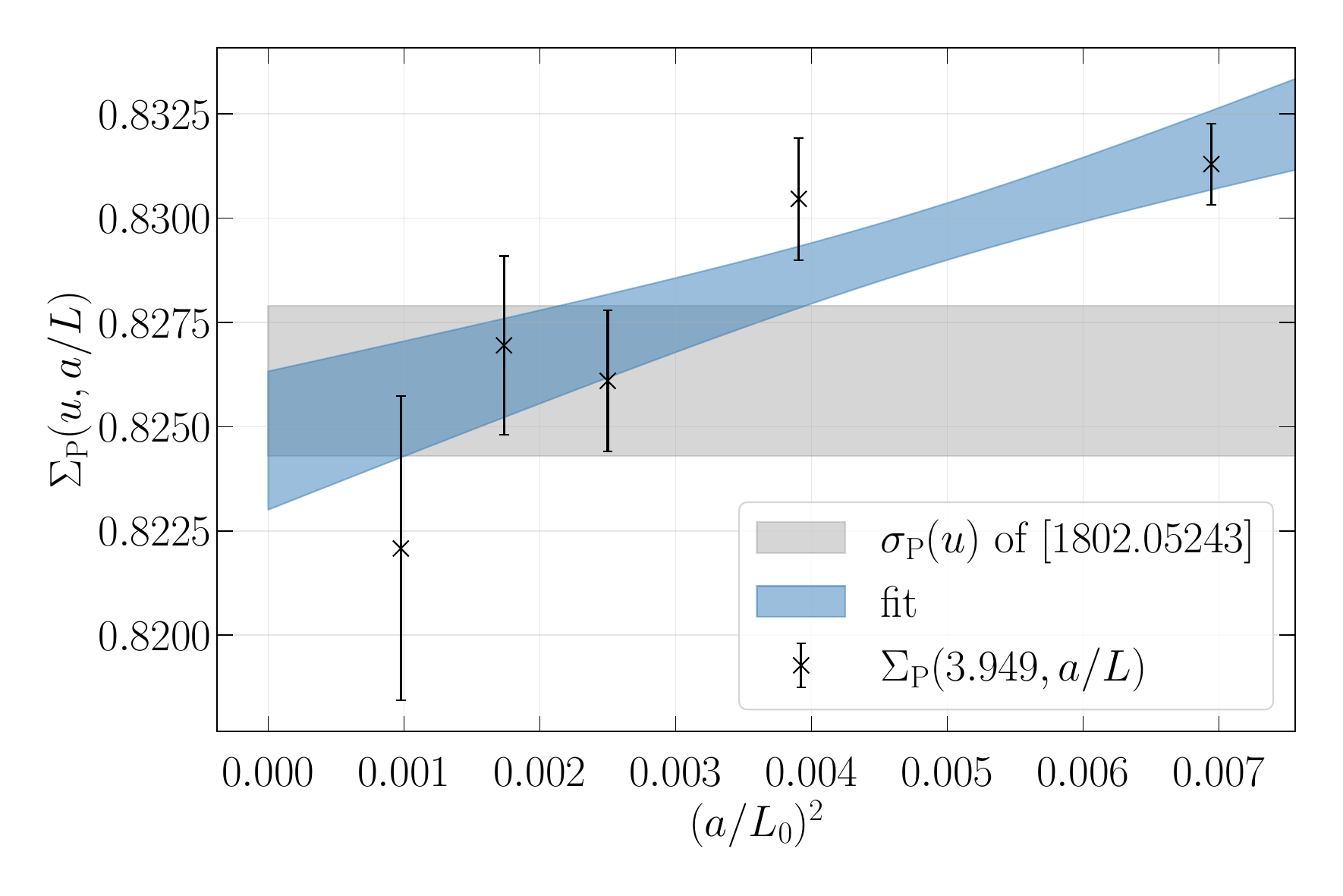}%
	\includegraphics[width=.5\textwidth]{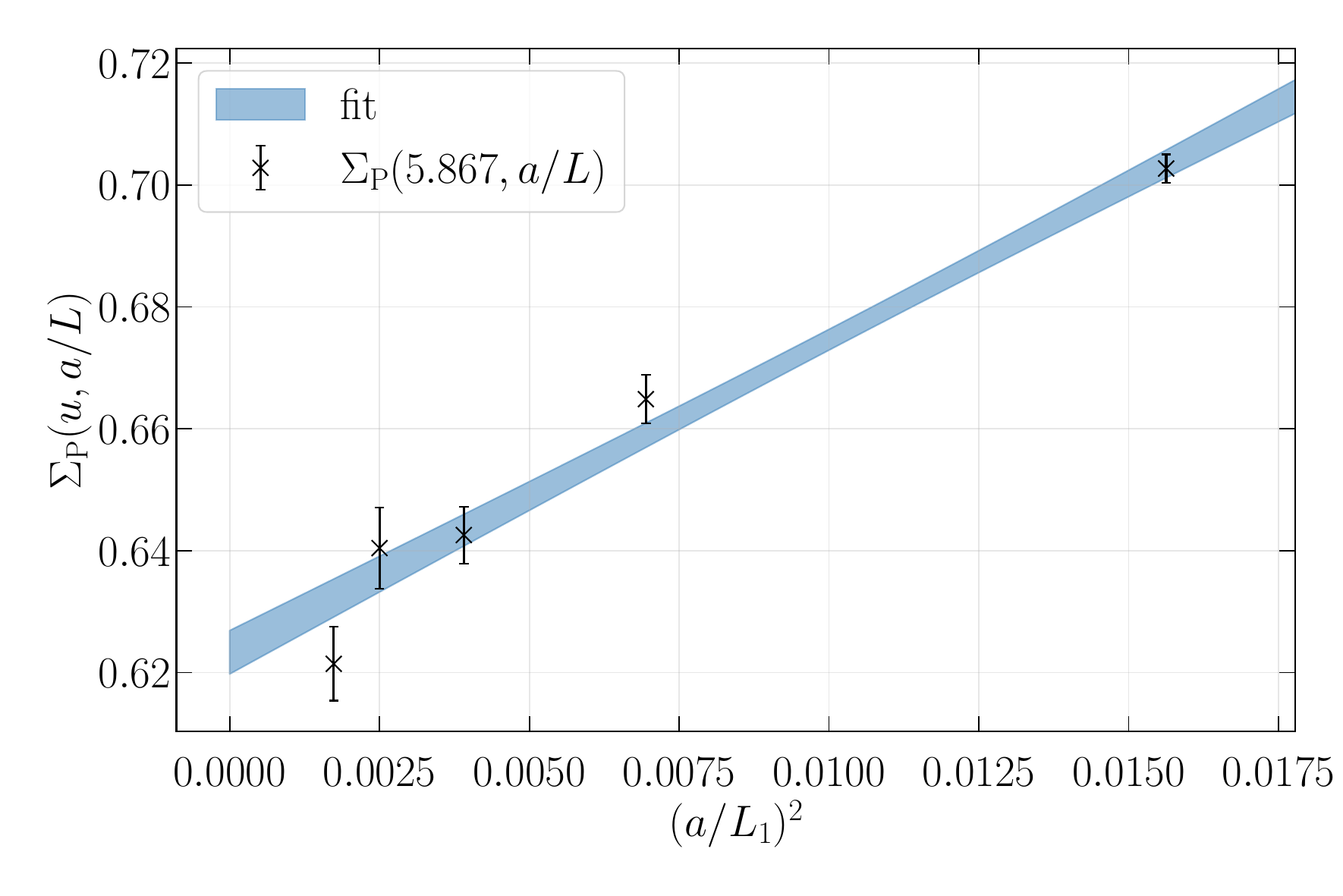}\\%
    \caption{Step-scaling function of the mass, measured on the lattices with
             the bare parameters fixed in $L_0$ (left) and $L_1$ (right). The
             gray error band shows the  global fit to these step-scaling
             functions performed in ref.~\cite{Campos:2018ahf}. Again, this fit
             did not include the data points presented here.
	        }
	\label{fig:ZPstep}
\end{figure}

The step-scaling function of the quark mass is identical to the (inverse of)
the step-scaling function of $\ZP$, see~\cite{Capitani:1998mq}. We show our
results in figure~\ref{fig:ZPstep}. For the step $L_0\to L_1$, we can again
compare directly to previous estimates~\cite{Campos:2018ahf} for the same
lattice action, and the agreement is very good as shown on the l.h.s. of
figure~\ref{fig:ZPstep}. The step  $L_1\to L_2$, however, was not covered by
\cite{Campos:2018ahf}. Our data can be used to further constrain and check the
global fit obtained in~\cite{Campos:2018ahf} and hence reduce the uncertainty
of the non-perturbative renormalisation of the quark mass. 

\subsection{Improvement of the axial current revisited}\label{sec:ca}
\begin{figure}
	\centering
	\includegraphics[height=.3\textheight]{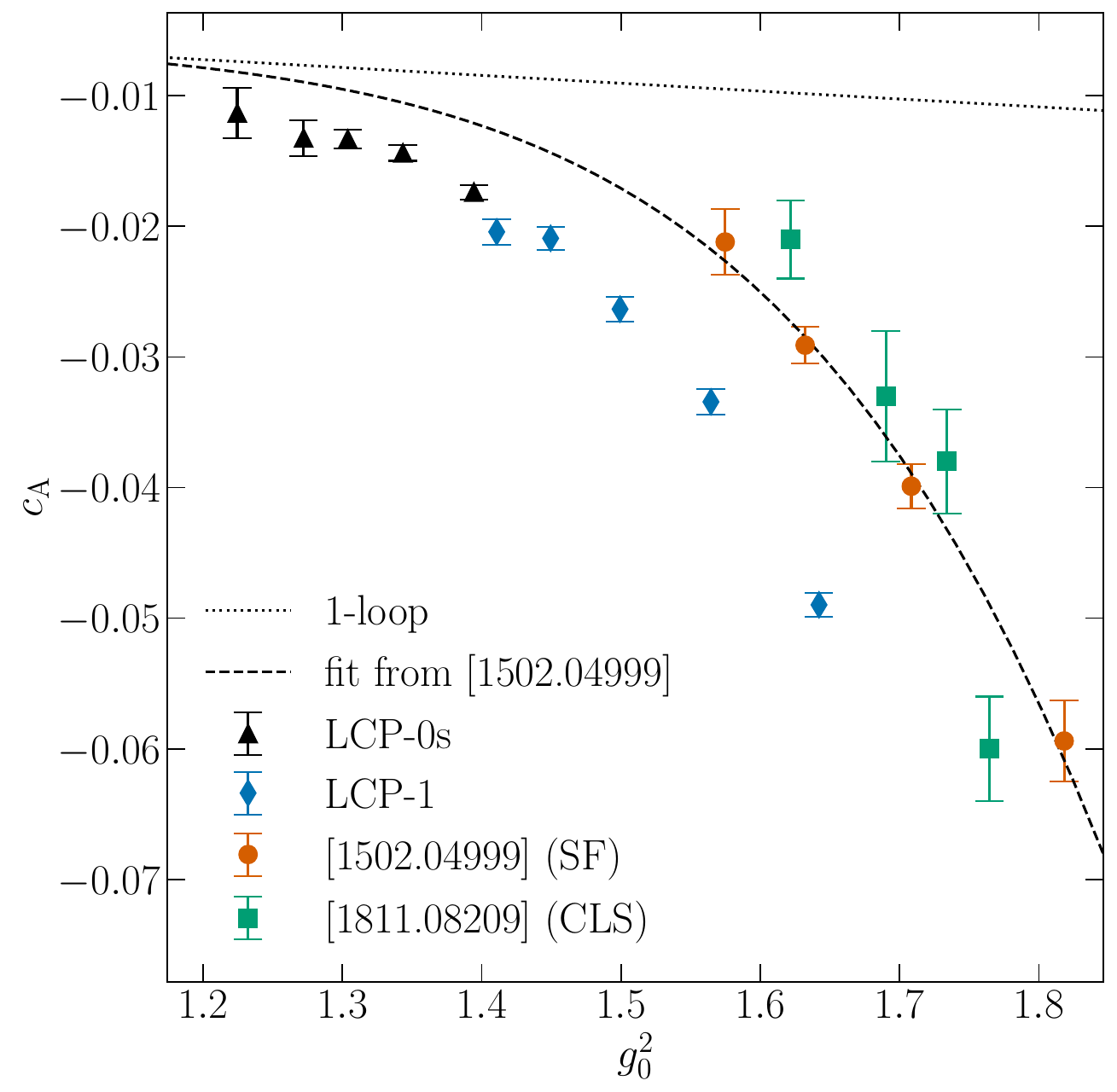}%
	\hfill
	\includegraphics[height=.3\textheight]{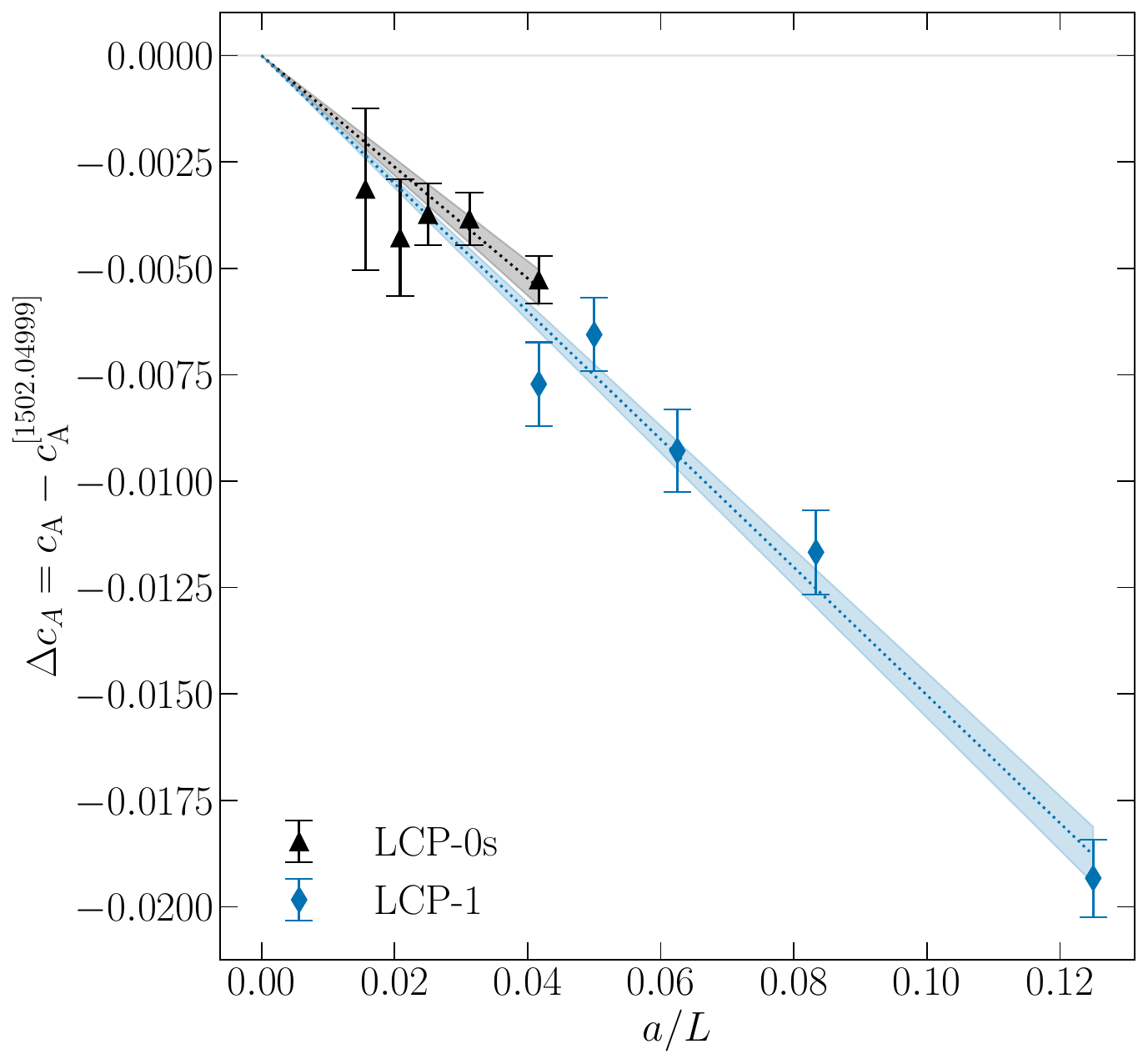}%
    \caption{\textit{Left:} Improvement coefficient $\ca$ as determined on
             several lines of constant physics.  LCP-0s and LCP-1 denote the
             results obtained in this work, while SF refers to those
             of~\cite{Bulava:2015bxa}; points labelled CLS are from the
             large-volume calculation in~\cite{Gerardin:2018kpy}.  The dotted
             line indicates the one-loop prediction~\cite{Aoki:1998qd}, and the
             dashed line the fit result of~\cite{Bulava:2015bxa}. Note that both
             LCP-0s and LCP-1 are very close to each other: they have  $L\approx
             L_1$ and $m\approx0$.  \textit{Right:} Difference between our
             evaluation of $\ca$ and the fit from~\cite{Bulava:2015bxa} as a
             function of $a/L$ together with one-parameter fits to the data.
	        }
	\label{fig:cA}
\end{figure}
A non-perturbative interpolating fit formula for the $\rmO(a)$ improvement
coefficient of the axial-vector current, $\ca(g_0^2)$, was determined for our
three-flavour lattice QCD action in ref.~\cite{Bulava:2015bxa}.  As in the case
of $Z_\mathrm{A}$, it naturally applies to the parameter range where the
underlying data were generated, i.e., to the large-coupling region (resp.\ the
associated $\beta$-range) of state-of-the-art large-volume computations, while
here we work at very small lattice spacings which means significantly larger
$\beta$.  Therefore, this once again raises the question of the reliability of
the interpolation formula found in~\cite{Bulava:2015bxa} beyond its immediate
scope of applicability.

To clarify this question, we have calculated $\ca$ directly on our gauge field
ensembles, adopting an alternative improvement condition compared
to~\cite{Bulava:2015bxa}. As originally suggested in
ref.~\cite{Luscher:1996sc} as advantageous, it amounts to varying the (spatial)
periodicity angle $\theta$ of the fermion fields, viz.
\begin{eqnarray}
  \psi_i(x+\hat k L) = e^{\rmi\theta}\psi_i(x)\,,\quad \psibar_i(x+\hat k L) = \psibar_i(x)e^{-\rmi\theta}\,,\quad k=1,2,3\,.
\end{eqnarray}
More specifically, we add quenched quarks, mass-degenerate with the sea quarks
but with different $\theta$. Adding an argument $\theta$, we consider their
PCAC masses $m_{ij}(x_0,\theta)$ and parameterise their values at $x_0=T/2$ in
the form
\begin{eqnarray}
	m_{12}(T/2,\theta) 
	= r(\theta) + a \ca s(\theta)\,.
\end{eqnarray}
The improvement coefficient $\ca$ can then be estimated from
\begin{eqnarray}\label{eq:cAdef}
	\ca= -{r(\theta)-r(\theta') -\left[r^{(0)}(\theta) - r^{(0)}(\theta')\right]\over a(s(\theta)-s(\theta'))}\,,
\end{eqnarray}
where $r^{(0)}$ denotes the value at $g^2_0=0$ as listed in table~\ref{tab:cA_tl}.
In particular we consider $\theta=0.5$ and $\theta'=0$. We display the results
for $L\approx L_1$ and compile them as a function of $g_0^2$ alongside the fit
formula from \cite{Bulava:2015bxa} on the left hand side of
figure~\ref{fig:cA}. The comparison demonstrates that the fit formula is
accurate throughout the relevant range of $g_0^2$ up to the always present
$a/L$ ambiguity. This ambiguity is illustrated on the right hand side of the
figure, where we show the difference between our results and the interpolating
formula together with fits to a linear function in $a/L$.  We therefore use the
existing fit formula from ref.~\cite{Bulava:2015bxa} in this paper and
recommend it for future applications.

%% file: tables/enstab_L0.tex
\begin{tabular}{@{\extracolsep{-0.1cm}}cccccccrrc}\toprule
$L/a$ & $\beta$ & $\kappa$ & $\bar{g}^2_\mathrm{GF}(L)$ 
                                                & $L\,m$    & $\ZA$   & $\ZP(L)$ & $N_{\rm cfg}$ & $\dfrac{\tau_{\rm ms}}{\rm MD}$  \\\midrule 
  $12$  & $4.3030$ & $0.1359947$ & $3.9461(43)$ & $-0.00024(33)$ & $0.831822(67)$ & $0.57835(32)$      & 9669          &  8                               \\
  $16$  & $4.4662$ & $0.1355985$ & $3.9475(58)$ & $+0.00043(30)$ & $0.838846(61)$ & $0.56972(44)$      & 5887          &  10                              \\
  $20$	& $4.6017$ & $0.1352848$ & $3.9493(63)$ & $+0.00100(20)$ & $0.844881(35)$ & $0.56503(50)$      & 8478          &  10                              \\                                               
  $24$  & $4.7165$ & $0.1350181$ & $3.9492(62)$ & $+0.00012(17)$ & $0.849684(29)$ & $0.56002(47)$      & 7303          &  16                              \\                                               
  $32$  & $4.9000$ & $0.1345991$ & $3.9490(97)$  & $+0.00562(15)$ & $0.857011(26)$ & $0.55388(69)$      & 5014          &  20                              \\\bottomrule
\end{tabular}

%% file: tables/enstab_2L0.tex
\begin{tabular}{@{\extracolsep{0.0cm}}cccccrc}\toprule
    $L/a$ & $\beta$  & $\kappa$ & $\bar{g}^2_\mathrm{GF}(L)$ 
                                                & $(L/2) \cdot m$     & $N_{\rm cfg}$
                                                                       & $\dfrac{\tau_{\rm ms}}{\rm MD}$    \\ \midrule
	24      & 4.3030   & 0.1359947  & 5.592(11) & +0.00430(10) & 12000 &  10  \\
	32      & 4.4662   & 0.1355985  & 5.689(14) & +0.00380(7)  & 13500 &  10  \\
	40      & 4.6017   & 0.1352848  & 5.755(22) & +0.00402(6)  & 10047 &  10  \\
	48      & 4.7165   & 0.1350181  & 5.770(25) & +0.00305(8)  & 3190  &  30  \\
	64      & 4.9000   & 0.1345991  & 5.844(32) & +0.00742(7)  & 2502  &  40  \\ \bottomrule
\end{tabular}

%% file: s_renorm_heavy.tex
\section{Massive quenched flavours}\label{s:massive}

Quenched heavy quarks are essential for the planned B-physics programme. Again
the starting point is to fix the LCP, now including the bare heavy quark mass.
Furthermore, a first application will be the determination of the b-quark mass~\cite{Kuberski:2026bpx}, and we thus need to compute the
renormalisation factor connecting bare quark masses to the renormalisation
group invariant (RGI) quark masses (cf. section~\ref{sec:massive}).
Here we briefly review the renormalisation and improvement of the quark masses
in a standard mass-independent SF scheme and then also in partially massive
schemes, where some mass-enhanced $a^2$-effects are absorbed into
renormalisation factors and improvement coefficients.  A complete
mass-dependent scheme as in \cite{Fritzsch:2018kjg} would also require the
determination of mass-dependent coefficients $c_\mathrm{A}(g_0^2,a\mq)$ and
$c_\mathrm{sw}(g_0^2,a\mq)$. It is beyond the scope of our project.

We refer to ref.~\cite{deDivitiis:2019xla} (appendix A) for a detailed
discussion of the connection between the renormalisation of the unitary theory
and that of additional quenched quarks. 

\subsection{Quark mass renormalisation}

Wilson fermions break chiral symmetry of QCD. An additive mass-counterterm is
needed to restore it in the continuum limit. For any bare quark mass of
flavour $i=1, \ldots, \Nf$ in the lattice action, one defines the bare
subtracted quark mass
\begin{align}\label{eq:bare-mass} 
        \mqi \equiv m_{0,i} - m_{\rm crit}(g^2_0) 
\end{align}
with a linearly divergent counterterm, $m_{\rm crit}(g^2_0) \sim a^{-1} (k_1
g_0^2 +\ldots)$. The standard definition for $m_{\rm crit}(g^2_0)$ is that all
$\mqi$ vanish for vanishing PCAC masses $m_{ij}=0$, $i,j\in\{1,\dots,\nf\}$.
In terms of the subtracted masses $\mqi$, the renormalised, $\rmO(a)$ improved
quark mass is given by
\cite{Bochicchio:1985xa,Luscher:1996sc,Bhattacharya:2005rb}
\begin{align}  \label{eq:mren-mq} 
        m_{i,\rm R} &= \zm \left\{ \left [\, \mqi  + (r_{\rm m} - 1){\Tr[\Mq]}/{\Nf} \,\right ] + a B_{i} \right\}  +  \rmO(a^2) \,,  \\\notag
        B_{i} &= \bm \mqi^2 + \bar b_{\rm m} \mqi \Tr[\Mq] + (r_{\rm m} d_{\rm m} - \bm) {\Tr[\Mq^2]}/{\Nf}  +  (r_{\rm m} \bar d_{\rm m} - \bar b_{\rm m}) {\Tr[\Mq]^2}/{\Nf}  \,,
\end{align}   
where $\Mq = {\rm diag}(m_{{\rm q},1}, \ldots , m_{{\rm q},\Nf})$ is the bare
subtracted mass matrix of the sea quarks, and $B_{i}$ is a combination of
Symanzik counterterms cancelling $\rmO(a)$ mass-dependent cutoff effects. $\zm$
depends on a renormalisation scale $\mu$, but all other coefficients are only
functions of $g^2_0$ when lattice spacing effects can be neglected. 

As already alluded to above, one can also renormalise and $\rmO(a)$ improve the
current (aka PCAC) quark masses, 
\begin{align}
    m_{ij,\rm R} &= \dfrac{\ZA}{\ZP}
                    \dfrac{1+\ba\,a\mqq[,ij] + \bar b_{\rm A}\,a \Tr[\Mq]}%
                          {1+\bp\,a\mqq[,ij] + \bar b_{\rm P}\,a \Tr[\Mq]}%
                          \cdot m_{ij}  +  \rmO(a^2) \,,
        \label{eq:mijR}
\end{align}
with $\mqq[,ij]=(\mqq[,i]+\mqq[,j])/2$. The explicit appearance of the quark
mass matrix in such equations reflects the fact that a mass-independent
renormalisation scheme has been chosen. Accordingly, all renormalisation
factors and improvement coefficients are independent of the masses in the
action at leading order in $a$, while improvement terms involve positive
powers of the mass matrix.%
\footnote{For completeness we mention that in the $\rmO(a)$ improved theory,
        and in a mass-independent renormalisation scheme, varying masses at
        fixed lattice spacing  requires keeping  $\tilde g_0^2 = g_0^2 (1+
        \bg(g_0^2) a \Tr[\Mq])$ fixed. Renormalisation factors such as
        $\za,\zp$ are therefore taken to depend on $\tilde g_0^2$. This
        mass-dependence is  relevant for heavy sea quarks as, e.g., used in
        \cite{Brida:2025gii}, while we here have $\Tr[\Mq]=0$.
}

We now consider massive valence quarks in our target $\Mq=0$ theory where
eq.~\eqref{eq:mren-mq} simplifies to
\begin{align}  \label{eq:mren-mq-simple}
  m_{i,\rm R} &= \zm \left\{ \mqi  + a \bm \mqi^2 \right\}  +  \rmO(a^2) \,.
\end{align}
For the remaining improvement coefficients needed in the quenched sector, we
follow standard practices and determine them by requiring the continuum PCAC
relation to be satisfied up to $ \rmO(a^2)$,
\begin{eqnarray}\label{eq:mijR2}
	m_{ij,\rm R} \stackrel{!}{=} \frac12 \left[ m_{i,\rm R}+m_{j,\rm R}\right] + \rmO(a^2)\,.
\end{eqnarray}
We explicitly employ the procedure originally developed in
ref.~\cite{Guagnelli:2000jw} with the improvements suggested in
ref.~\cite{deDivitiis:2019xla}. Imposing \eqref{eq:mijR2},
eqs.~\eqref{eq:mren-mq-simple} and \eqref{eq:mijR} lead to the following
relation between the bare current and subtracted quark masses at fixed gauge
coupling $g_0^2$,
\begin{align}\label{eq:massmatch}
    m_{ij}  &=  Z \left\{  \mqn{ij} +  a\bm \frac{1}{2}({\mqn{i}^2+\mqn{j}^2})  -  a(\ba-\bp)\mqn{ij}^2 \right \}   +  \rmO(a^2) \,,  \\
    Z &=  {\zm(\mu) \zp(\mu)}/{\za}                            \,, \notag
\end{align} 
which allows to determine the set of coefficients $Z$, $\bm$ and
$\bAP\equiv\bA-\bP$ non-perturbatively.
Note that the $\mu$-dependence cancels in $\zm\zp$ and that one cannot
disentangle the coefficients $\bA$ and $\bP$, since only their difference
appears after expanding the denominator for small $a\mqq[,ij]$
in~\eqref{eq:mijR}.
Furthermore, higher-order mass effects can be ignored, as they contribute only
at order $a^2$.

As explained in detail in section~2 of ref.~\cite{deDivitiis:2019xla}, it is
convenient to consider the specific set of masses
\begin{align}
  \mqn{i}=\epsilon\,,\,\text{for}\; i\leq \nf=3\,,\quad \mqn{4}=\mqn{5}=2\Delta + \epsilon
  \,,\quad \mqn{6}=\mqn{7}=\Delta + \epsilon\,,
\end{align}
which depend on two mass-parameters $\Delta$ and $\epsilon$. For $\epsilon=0$,
we have three massless sea quarks and two doublets of degenerate valence quarks
with masses $2\Delta$ and $\Delta$, respectively.  Using eq.~\eqref{eq:massmatch}
for flavour combinations $(i,j)\in\{(1,2)$, $(1,4)$, $(4,5)$, $(6,7)\}$, one
constructs the explicit estimators
{
\begin{subequations}\label{eqs:RX-Delta}
\begin{align}\label{eq:estim-RZ}
    \RZ  &\equiv \dfrac{\mij[45]-\mij[12]}{2\Delta} + a \left(\RAP-\Rm\right)(\mij[45]+\mij[12]) \,, \\[0.4em]\label{eq:estim-Rm}
    \Rm  &\equiv \dfrac{2\left(\mij[14]-\mij[67]\right)} {(\mij[45]-\mij[12])\,a\Delta}          \,, \\[0.4em]\label{eq:estim-RAP}
    \RAP &\equiv \dfrac{2\mij[14]-\mij[45]-\mij[12]}{(\mij[45]-\mij[12])\,a\Delta} \,.  
\end{align}
\end{subequations}
They determine $Z$ up to corrections of order $a^2$, and $\bm$ and $\bAP$ up to
$\rmO(a)$, respectively.  The mix of non-degenerate ($\mij[14]$) and degenerate
($\mij[45],\mij[67]$) PCAC masses is needed to separate $\bm$ from $\bAP$.
The estimators are also valid for any finite $\epsilon=\rmO(\mij[12])$. We
utilise them only with small values of $\epsilon$ to account for our
mistunings of the sea quark masses, i.e., for the fact that they are not exactly zero.

Taking the limit $\Delta \to 0$ defines our mass-independent scheme for  $Z$
and the two $b$-parameters. 
The so-defined functions $Z(\tilde g_0^2)$, $\bm(\tilde g_0^2)$ and
$\bAP(\tilde g_0^2)$ still have a residual dependence on the details of the
definitions of the PCAC masses, 
namely, on our choice of an SF scheme with specific angles $\theta$ and
classically-improved finite difference operators as in 
refs.~\cite{deDivitiis:1997ka,Guagnelli:2000jw,Heitger:2003ue,Fritzsch:2010aw,deDivitiis:2019xla}
which are advantageous for heavy quarks.
These
dependences are $\rmO(a^2)$ for $Z$ and $\rmO(a)$ for the $b$-parameters and
change only $\rmO(a^2)$ effects in the improved theory.
Below we define what we call a partially massive scheme (or ``massive scheme''
for short), which absorbs some of the $(a\mq)^n$-effects, $n>1$, into
mass-dependent $Z, \, \bm,\, \bAP$ and evaluate them numerically.
This scheme is similar to the one in
\cite{deDivitiis:2019xla,Fritzsch:2010aw,Heitger:2003ue} and goes part of the
way to a completely massive scheme whose advantages have been discussed
in~\cite{Fritzsch:2018kjg}.  

\subsection{Partially massive scheme}\label{sec:massive}

It is not necessary to take the limit $\Delta\to0$ in the above estimators.
Instead it is expected to be advantageous to choose $\Delta$ to be of the order
of the (bare) quark mass under consideration. This then introduces higher
orders in $a\mq$, which will absorb some of the higher-order cutoff effects.
Such a definition of a massive scheme is not complete, and mass-dependent
cutoff effects will remain, since additional improvement terms  with a
different structure are required. Let us just give one example.  Beyond the
first order in $a\mq$,  terms $a^2 \mqq[,i]\mqq[,j]$ are needed in numerator and
denominator of eq.~\eqref{eq:mijR}. These are not absorbed by mass-dependent
$Z, \, \bm,\, \bAP$. Nevertheless it has been observed in
\cite{Fritzsch:2010aw,Heitger:2003ue} that taking $\Delta$ around the quark
mass of interest  reduces cutoff effects compared to $\Delta=0$. 

For the practical choice of $\Delta$, one needs to add a condition keeping the
heavy {\em renormalised} quark mass fixed and determine $a\Delta$ for each
$a/L$ on the desired  LCP. We are interested in extending LCP-0,
table~\ref{tab:ens_L0}, in this way. 

\input{massive_determination.tex}

%% file: massive_determination.tex
As in
\cite{Capitani:1998mq,DellaMorte:2005kg,Campos:2018ahf,Fritzsch:2010aw,Fritzsch:2012wq,Bernardoni:2013xba}
we fix the RGI heavy quark mass, $\Mh$, which is scale and scheme
independent.%
\footnote{Here, the symbol $M$ is not to be confused with the matrix of bare
          subtracted sea quark masses appearing in the foregoing subsection.
}
The dimensionless variable
\begin{align}\label{eq:def-zh}
        \zh &= L_0\Mh  \,,
\end{align}
then has to remain constant as $a/L_0$ is varied.  In terms of the
subtracted
quark mass $m_{\rm h,R}(\mu)$, renormalised at scale $\mu$, the $z$-variable
reads
\begin{align}\label{eq:zh-lim}
        \zh &= L_0 \, h(\mu) \, m_{\rm h,R}(\mu)  \notag\\ 
            &= L_0 \, h(\mu) \,  \dfrac{\ZA(g_0^2)}{\ZP(g_0^2,a\mu)}  Z(g_0^2,a\Del[h]) \left\{ 1+a\bm(g_0^2,a\Del[h]) \mqh \right\} \mqh  \,, 
\end{align}
where the flavour-independent function $h(\mu)=M/m_{\rm R}(\mu)$ connects the
PCAC quark mass renormalised at scale $\mu$ to its renormalisation group
invariant counterpart. For the three-flavour theory, this factor has been computed
non-perturbatively in the continuum limit in ref.~\cite{Campos:2018ahf}. At
$\mu=1/L_0$ it evaluates to
\begin{align}\label{eq:hL0}
        h(1/L_0) &\equiv \left.\dfrac{M}{m_\mathrm{R}(\mu)}\right|_{\mu=1/L_0}
                       = 1.4744(87)  \,. 
\end{align}

To fix $\zh$ to a prescribed value at finite lattice spacing, we still need $Z$
and $\bm$ on each lattice as a function of the mass parameter $\Del$, and to
find the point $\Del=\Del[h]$ that corresponds to $\zh$.
\begin{figure}[t!]
        \centering\small
        \includegraphics[height=0.33\textheight,page=1]{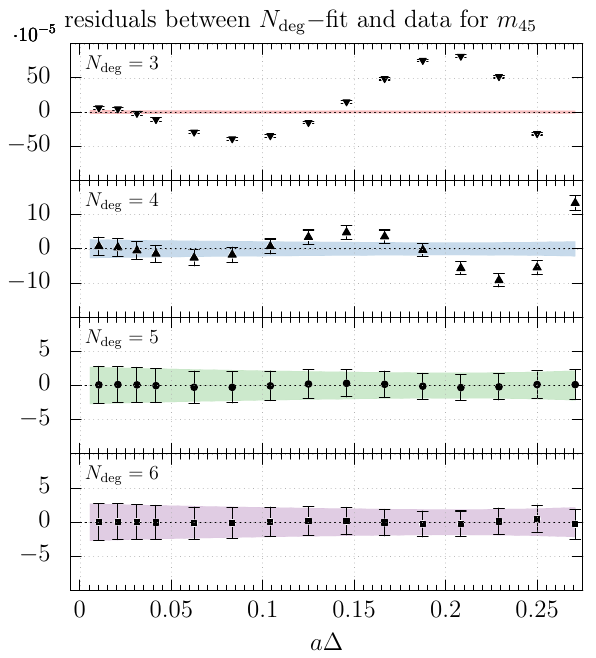}\hfill
        \includegraphics[height=0.33\textheight,page=2]{plots/mijfit_plot.pdf}
        \caption{PCAC masses and polynomial representation for ensemble $L_0/a=12$ (LCP-0).
                \textit{Left:}  Difference $a\mij[45] - a\mij[45]^\mathrm{fit}(\Delta)$ between the measured
                PCAC masses $m_{45}$ and their fit $a\mij[45]^\mathrm{fit}(\Delta)$ for a
                given polynomial degree $N_{\rm deg}$.  Statistical
                uncertainties are those of the data points and fit,
                respectively. \textit{Right:} Global fit to
                eq.~\eqref{eq:mass_series} with $N_{\rm deg}=5$. Uncertainties
                are too small to show up on this scale.
                Note that $2a\Delta\approx a\mqq[,4]$, such that data on 
                $L_0/a=12$ extends up to $a\mqq[,4]\approx 0.5$.
            }
        \label{fig:n_fit}
\end{figure}
To map out the relevant heavy  quark sector numerically, we select quark
masses to generously cover the charm and bottom quark mass regions, i.e., in
the range $0.25 < L_0 \mqn{4} \leq 6.5$. For $\mathrm{O}(15)$ quark masses we
compute the current quark masses $m_{45}$ and $m_{14}$, respectively. In the
spirit of ref.~\cite{deDivitiis:2019xla}, we parametrise the calculated current
quark masses by polynomials of degree $n$ according to
\begin{subequations} \label{eq:mass_series}
\begin{align}
        am_{14}^\mathrm{fit}(\Delta) &= am_{12}(0) + \phantom{2}N_1 a \Delta + \phantom{2}N_2(a\Delta)^2 + \dots + \phantom{2}N_n(a \Delta )^n \,, \label{eq:m12_series}\\
        am_{45}^\mathrm{fit}(\Delta) &= am_{12}(0) +2 N_1 a \Delta + D_2(2a\Delta)^2 + \dots + D_n(2 a \Delta )^n                     \,. \label{eq:m22_series}
\end{align}
\end{subequations}
Systematic tests confirm that these current quark masses are best represented
by polynomials of degree $n=5$ in our range of data.  We perform a simultaneous
fit of the two ansätze in eq.~\eqref{eq:mass_series}, where $am_{12}(0)$ is
exactly fixed to the measured value of the sea quark mass, cf.
table~\ref{tab:ens_L0}, resulting in 9 fit parameters for $\mathrm{O}(30)$
correlated data points. We show a representative fit result together with the
input data in the right panel of figure~\ref{fig:n_fit}. The left panel shows
the quality to which the data can be
represented by a polynomial of given degree.
These ans{\"a}tze smoothly connect the valence sector of heavy quarks to the
massless unitary point and provide a representation of the data that allows us
to evaluate eqs.~\eqref{eqs:RX-Delta} for any valence mass in the range covered
by data. In our setup with $\Nf=3$ degenerate quarks, the valence mass
parameterising the heavy quark mass dependence is given by
\begin{table}[t!]
	\setlength{\abovecaptionskip}{2pt}
	\setlength{\belowcaptionskip}{10pt}
	\centering\small
	\renewcommand{\arraystretch}{1.25}
	\caption{
		Hopping parameters $\kappa_4(\zh,L_0/a)$ for a heavy-valence
		line of constant physics fixed by the dimensionless RGI heavy quark
		mass $z_\mathrm{h}=L_0M_\mathrm{h}$ determined from
		eq.~\eqref{eq:zh-lim} on the LCP-0 ensembles.
		Choices of $\zh$ that have been used for measurements in the context of~\cite{Conigli:2023rod,Kuberski:2026bpx} are denoted by an asterisk.
	}
	\label{tab:kappas}
	\input{./tables/tab_kappas_matching.tex}

\end{table}
\begin{align}\label{eq:Delta}
        a\Delta &= \dfrac{a}{2}\Big[\mqq[,4]-\mqq[,1]\Big] = \dfrac{1}{2}\Big[ ({2\kappa_{4}})^{-1}-({2\kappa_{1}})^{-1} \Big] \,.
\end{align}
This parameter is independent of the critical mass or hopping parameter.
In our specific case ($\mqq[,1]\approx 0$) we have $\mqq[,\rm 4] \approx 2\Delta$.
Having collected all data, we plug the polynomial representations
(\ref{eq:m12_series}) and (\ref{eq:m22_series})
into eq.~\eqref{eq:zh-lim} and
impose target values for the dimensionless heavy RGI mass $\zh$ to solve for
the bare parameters $\Delta=\Del[h]$ and thus $\kappa_4 =
(4\Delta_{\mathrm{h}} + 1/\kappa_{1})^{-1}$. The so obtained hopping parameters
$\kappa_4(\zh,L_0/a)$ are presented in table~\ref{tab:kappas}
and have been used in \cite{Conigli:2023rod,Kuberski:2026bpx}.
Each row represents a single heavy-valence line of constant physics for the
light LCP-0 of table~\ref{tab:ens_L0}.  The values of $\zh$ denoted by an
asterisk, encompassing the charm and bottom regions, have been used to define
LCPs for the measurement of heavy-quark observables in the context of the
applications in refs.~\cite{Conigli:2023trw,Conigli:2023rod} and the
non-perturbative matching of HQET and QCD along the lines of
ref.~\cite{DellaMorte:2013ega}.
The central values and full covariance matrices of the fit parameters entering
eqs.~(\ref{eq:m12_series}) and (\ref{eq:m22_series}), together with $\ZA$ and
$\ZP$ from table~\ref{tab:ens_L0}, are available in the accompanying Zenodo
record~\cite{fritzsch_2026_20554739}. They can be used to reproduce the
results in table~\ref{tab:kappas}, including the propagation of correlated
uncertainties.

\begin{figure}[t!]
    \centering\small
    \includegraphics[width=\linewidth]{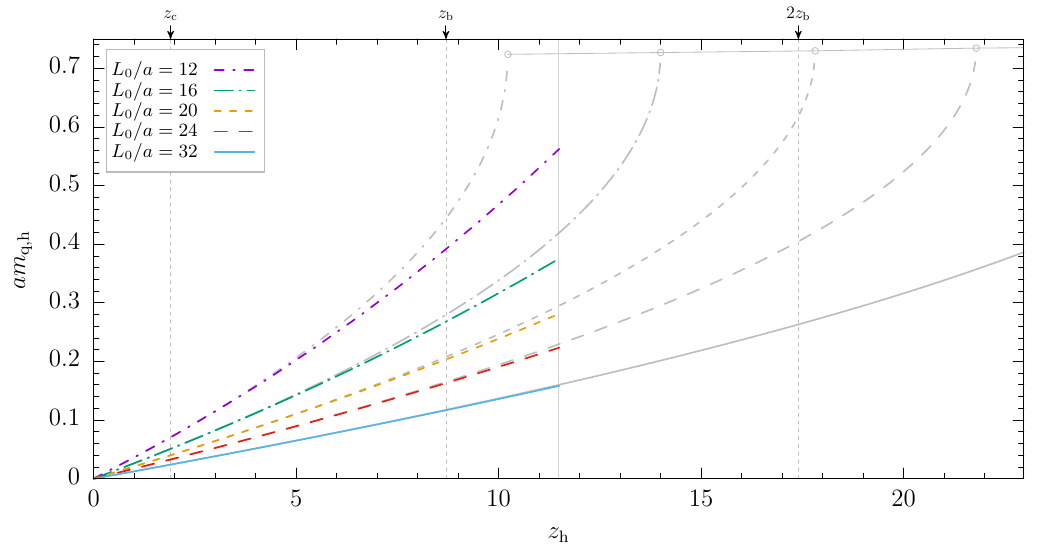}
    \caption{Subtracted bare quark masses corresponding to valence
             hopping parameters $\kappa_4(\zh,L_0/a)$.
             The gray curves are
             determinations in the chiral, unitary theory ($\Delta=0$)
             obtained by inverting the quadratic eq.~\eqref{eq:zh-lim} with
             estimators $\RX(g_0^2,0)$. The (connected) gray circles are
             the end points in the massless scheme and hardly vary with the
             lattice spacing. The coloured curves are determinations in our 
             massive scheme (varying $\Delta$).  The vertical gray line
             in the centre represents our reach in valence masses for which
             we accumulated data to determine the estimators $\RX$ (up to
             approx.\ $z=1.32 \zb$). The grey dashed vertical lines indicate
             the charm ($\zc=1.9$), bottom ($\zb=8.7$) and twice the bottom
             quark mass.
            }\label{fig:kaph_vs_zh}
\end{figure}

We end this subsection with a comparison of the functional $\zh$-dependence of
the hopping parameters in our massive scheme (corresponding to
table~\ref{tab:kappas}) and the massless one.  The latter results are presented as gray
curves in figure~\ref{fig:kaph_vs_zh} and (as explained in
ref.~\cite{deDivitiis:2019xla}) obtained by restricting the calculation of the
estimators~\eqref{eqs:RX-Delta} to the first two parameters in
eqs.~\eqref{eq:mass_series}. 
One clearly observes end-points (open circles) for each lattice size, because
eq.~\eqref{eq:zh-lim} reduces to a quadratic equation in $\mqq[\rm, h]$ with
solutions in a limited domain of $z_\mathrm{h}$. The same applies at other a
priori fixed values $\Delta$ (as done in
refs.~\cite{Guagnelli:2000jw,Heitger:2003ue,Fritzsch:2010aw}). Such a choice
just moves the end-point towards higher values of the heavy-quark mass, but at
some point one enters a region where mass-dependent cutoff effects reappear.

Therefore, a well-motivated approach is to rather consider $\Delta$ as a
\emph{free} parameter until matched to a continuum value $\zh$ according to our
prescription described in the previous paragraph. It requires mapping out the
dependence of $Z$ and $\bX$ on $a\Delta$. Then, by imposing some fixed
numerical value for $\zh$, the corresponding value $\Delta\equiv\Del[h]$
follows. This translates into results for $\mqh$, presented as coloured curves
in figure~\ref{fig:kaph_vs_zh}, and listed as $\kappa_4(\zh,L_0/a)$ in
table~\ref{tab:kappas} for some specific input masses $\zh$.  They do not show
any problematic behaviour as there is no explicit truncation of higher-order
terms. The curves just end, because our data set is limited to masses
$\zh\lesssim 1.3\zb$.

\subsection{An illustration of the effectiveness of the partially massive scheme}\label{sec:tests-L0}

\begin{table}[t!]
	\setlength{\abovecaptionskip}{5pt}
	\setlength{\belowcaptionskip}{2pt}
	\centering\small
	\renewcommand{\arraystretch}{1.25}
    \caption{Estimators $\RX(\zh,L_0/a)$ along different valence lines of
            constant physics (based on LCP-0) fixed either in our massive
            scheme ($\Delta=\Del[h]$) or in the massless scheme ($\Delta=0$).
            Both practically employ the numerical procedure described above,
            exploiting the matching condition for
            heavy quark masses, eq.~\eqref{eq:zh-lim}, together with 
            the input values $\zc$ and $\zb$ from eq.~\eqref{eq:zh-fixed}.
        }
	\input{./tables/RX_all.tex}
	\label{tab:RXall}
\end{table}
\begin{table}[t!]
	\setlength{\abovecaptionskip}{10pt}
	\setlength{\belowcaptionskip}{0pt}
	\centering\small
	\renewcommand{\arraystretch}{1.25}
    \caption{Hopping parameters $\kappa_4(\zh,L_0/a)$ for different valence
             lines of constant physics, defined as in table~\ref{tab:RXall}.
            }
    \input{./tables/tab_kappas_phys.tex}
	\label{tab:kappas_phys_z}
\end{table}

We present one illustration  of the massive scheme on the LCP-0. The heavy
quark mass is fixed by imposing a specific value on $\zh$,
eq.~\eqref{eq:zh-lim}. For the charm quark we set
$\Mc=1486(21)\,\MeV$~\cite{Heitger:2021apz}. For our illustration, the charm
and bottom quark masses need to correspond only approximately to their physical
values.
Thus, we choose the scale-independent ratio $\Mb/\Mc\approx 4.6$ to estimate
$\Mb$, by which we arrive at approximate values that we adopt for the
dimensionless RGI charm and bottom quark mass in $L_0$ here:
\begin{align}\label{eq:zh-fixed}
  \zc &= 1.9 \,, & \zb &= 8.7 \,.
\end{align}
Then, imposing those values onto $\zh$ on the different lattices $L_0/a$ as
explained above, and using either our new massive scheme ($\Delta=\Del[h]$) or
a massless scheme ($\Delta=0$), we obtain the estimators $\RX$ and hopping
parameters listed in tables~\ref{tab:RXall} and \ref{tab:kappas_phys_z},
respectively. At fixed heavy quark mass, the difference of the hopping
parameters between the two schemes corresponds to the difference of the curves
in figure~\ref{fig:kaph_vs_zh} at the vertical dashed lines for $\zh=\zb,\zc$.
We note right away that the hopping parameters obtained for the charm quark in
the two schemes essentially agree and no significant difference can be expected
in our range of lattice spacings.  However, with lattice spacings 4--5 times
larger, as typically used in large-volume simulations, the charm would naively
take the role of the bottom quark we are studying in small volume.  According
to table~\ref{tab:kappas_phys_z}, there is a significant difference in the
hopping parameters for the bottom quark, which will be our main focus in the following.
%

\subsubsection*{Scaling of current quark masses involving bottom quarks}

\begin{figure}[t!]
 \small
 \centering
 \includegraphics[width=\textwidth]{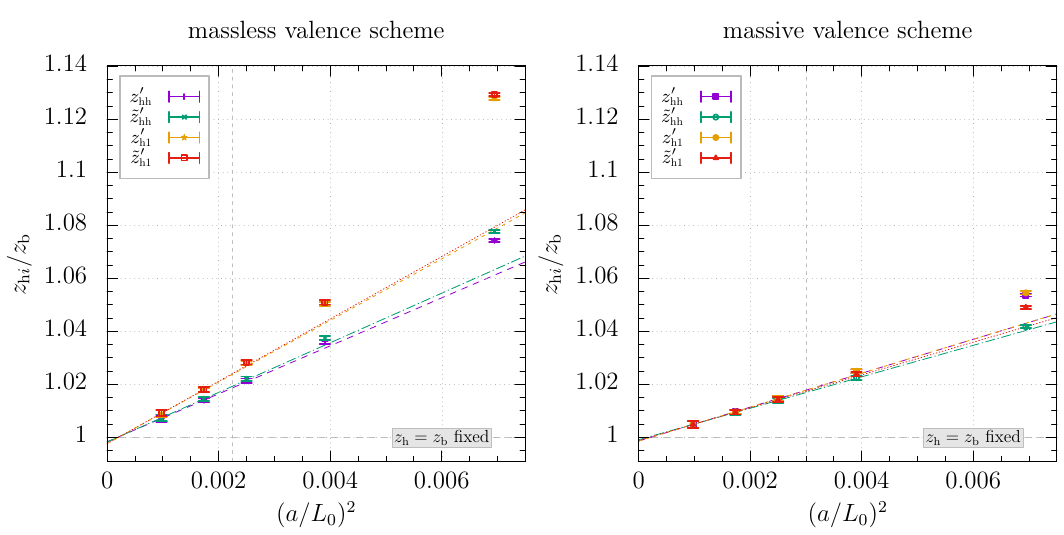}
 \caption{Lattice spacing dependence of various improved and renormalised PCAC
          masses (on LCP-0 ensembles) after fixing the RGI heavy quark mass from the
          subtracted quark mass, $\zh=\zb$, cf. eq.~\eqref{eq:zh-lim}.  The
          left panel shows results in the massless scheme and the right panel
          shows data in the massive scheme.  The data are normalised w.r.t. the fixed
          quark mass $\zb=8.7$ to better identify the size of cutoff effects.
          The extrapolating curves represent individual fits of the form
          $b_0+b_1(a/L_0)^2$ including data points to the left of the dashed
          vertical lines. 
         }\label{fig:L0-pcac}
\end{figure}

The estimators $\RX$, relevant for mass-improvement, have been derived from the
equality of (renormalised and improved) current and subtracted quark masses,
cf.\ eq.~\eqref{eq:massmatch}.  While the latter is used to fix the heavy
LCP, we can ask the question for residual cutoff effects on current quark
masses. The corresponding RGI current quark masses are defined as
\begin{align}\label{eq:def-zij}
  \zij &= \dfrac{L}{2}(M_{i}+M_{j}) 
        = h(\mu)\lim_{a\to 0}\left[ \dfrac{\ZA(g_0^2)}{\ZP(g_0^2,a\mu)}(1+\bAP(g_0^2,a\Delta) a\mqq[,ij])L\mij \right] 
        \,.
\end{align}
Using eq.~\eqref{eq:def-zij}, we define
\begin{align}\label{eq:def-zhi}
        z'_{\mathrm{hh}} &=  \zij[\mathrm{hh}]\,, &
        z'_{\mathrm{h1}} &= 2\zij[\mathrm{h1}] - \zij[{\rm 12}]\,.
\end{align}
At finite lattice spacing, both choices provide an estimator for the RGI mass
of the heavy quark flavour that should agree in the continuum with the value of
$\zh$ used to set up the heavy LCP. Moreover, we can employ the leading-order
relationship $Z\mqq[,ij]=\mij$ to replace the mass-improvement term $\bAP
a\mqq[,ij]$ through $\bAPt a\mij$, where $\bAPt=\bAP/Z$ with $Z$ taken in
the corresponding scheme. This provides two
additional estimators for the heavy quark mass which we denote as
$\tilde{z}'_{{\mathrm{h}i}}$. Our results for these four estimators along LCP-0
($L=L_0$) are presented in figure~\ref{fig:L0-pcac} for the massless and
massive scheme (left and right panel). All results are normalised to the
defining value of the heavy LCP, $\zb=8.7$, to easily read off the size of
cutoff effects.
The curves show simple linear extrapolations in $(a/L_0)^2$ for each data set,
using only data points left of the vertical dashed line.  While one observes
all extrapolations to lead to 1 in the continuum limit within uncertainties,
the massive scheme exhibits a significant reduction of cutoff effects for all
estimators compared to the corresponding results in the massless scheme.
As a result of the extended $(a/L_0)^2$-scaling range for the massive scheme,
the extrapolated value is also more precise. 

We  emphasise another key difference between the two schemes.  In the massive
scheme, we observe that $z_{\mathrm{hh}}'\approx z_{\mathrm{h1}}'$ at finite
lattice spacing.
This is not the case in the massless scheme, where the estimators for the heavy
RGI quark mass via heavy-light and heavy-heavy channels are clearly separated.

%% file: tables/tab_kappas_matching.tex
%
\begin{tabular}{llllll}\toprule
$\zh$  & $\kappa_4(z_\mathrm{h}, 12)$ 
                        & $\kappa_4(z_\mathrm{h}, 16)$ 
                                           & $\kappa_4(z_\mathrm{h}, 20)$ 
                                                              & $\kappa_4(z_\mathrm{h}, 24)$ 
                                                                                & $\kappa_4(z_\mathrm{h}, 32)$ \\ \midrule
 $ 1.0$ &$0.13466612(74)$&$0.13464165(80)$&$0.13452873(68)$&$0.13439940(53)$&$0.13414644(56)$\\
 $ 1.5$ &$0.1339864(11) $&$0.1341509(12) $&$0.1341458(10) $&$0.13408678(80)$&$0.13391837(85)$\\
 $ 1.75^{\ast}$ &$0.1336425(13) $&$0.1339034(14) $&$0.1339530(12) $&$0.13392964(93) $&$0.1338039(10)$\\
 $ 2.0$ &$0.1332958(15) $&$0.1336545(16) $&$0.1337594(14) $&$0.1337719(11) $&$0.1336891(11) $\\
 $ 2.125^{\ast}$ &$0.1331214(16) $&$0.1335295(17) $&$0.1336623(15) $&$0.1336929(11)  $&$0.1336316(12)  $\\
 $ 2.5$ &$0.1325940(19) $&$0.1331525(20) $&$0.1333696(17) $&$0.1334548(13) $&$0.1334587(14) $\\
 $ 3.0$ &$0.1318807(24) $&$0.1326446(24) $&$0.1329763(21) $&$0.1331355(16) $&$0.1332271(17) $\\
 $ 3.5$ &$0.1311556(28) $&$0.1321309(28) $&$0.1325794(25) $&$0.1328138(19) $&$0.1329943(20) $\\
 $ 4.0^{\ast}$ &$0.1304187(33) $&$0.1316111(33) $&$0.1321790(29) $&$0.1324897(22) $&$0.1327602(23) $\\
 $ 4.5$ &$0.1296696(38) $&$0.1310853(37) $&$0.1317748(32) $&$0.1321633(25) $&$0.1325250(26) $\\
 $ 5.5$ &$0.1281354(47) $&$0.1300149(46) $&$0.1309554(40) $&$0.1315031(31) $&$0.1320507(33) $\\
 $ 6.5$ &$0.1265532(58) $&$0.1289193(55) $&$0.1301209(49) $&$0.1308331(37) $&$0.1315714(39) $\\
 $ 7.5$ &$0.1249244(69) $&$0.1277982(65) $&$0.1292711(57) $&$0.1301531(44) $&$0.1310869(45) $\\
 $ 8.125^{\ast}$ &$0.1238834(76) $&$0.1270846(72) $&$0.1287322(63) $&$0.1297229(48)  $&$0.1307814(50) $ \\
 $ 8.375^{\ast}$ &$0.1234621(79) $&$0.1267964(74) $&$0.1285149(65) $&$0.1295497(49)  $&$0.1306586(51) $\\
 $ 8.5$ &$0.1232504(80) $&$0.1266517(76) $&$0.1284059(66) $&$0.1294628(50) $&$0.1305971(52) $\\
 $ 8.625^{\ast}$ &$0.1230380(82)  $&$0.1265066(77) $&$0.1282966(67) $&$0.1293758(51)  $&$0.1305355(53) $\\
 $ 8.875^{\ast}$ &$0.1226112(85) $&$0.1262153(80) $&$0.1280774(69) $&$0.1292013(53)  $&$0.1304121(55) $\\
 $ 9.5$ &$0.1215317(92) $&$0.1254802(87) $&$0.1275251(75) $&$0.1287623(57) $&$0.1301021(59) $\\
 $10.5^{\ast}$ &$0.119766(11)  $&$0.1242847(98) $&$0.1266289(84) $&$0.1280514(64) $&$0.1296017(66) $\\
 $11.5$ &$0.117948(12)  $&$0.123067(11)  $&$0.1257175(94) $&$0.1273304(71) $&$0.1290959(73) $\\
\bottomrule
\end{tabular}

%% file: tables/RX_all.tex
\begin{tabular}[]{cclllll}
	\toprule
	$X$ & $\zh$ & $R_X(\zh, 12)$  & $R_X(\zh, 16)$  & $R_X(\zh, 20)$  & $R_X(\zh, 24)$  & $R_X(\zh, 32)$  \\ \midrule
	AP  & $0$   & $-0.02580(21)$  & $-0.02951(74)$  & $-0.02800(21)$  & $-0.02848(25)$  & $-0.02763(29)$  \\
	AP  & $\zc$ & $-0.01289(16)$  & $-0.01802(28)$  & $-0.01933(15)$  & $-0.01996(17)$  & $-0.02052(20)$  \\
	AP  & $\zb$ & $+0.14771(26)$  & $+0.06119(18)$  & $+0.03083(12)$  & $+0.015986(94)$ & $+0.003342(88)$ \\
	\cmidrule(lr){1-7}
	$Z$ & $0$   & $+1.110600(57)$ & $+1.108992(58)$ & $+1.107285(32)$ & $+1.105309(29)$ & $+1.102069(25)$ \\
	$Z$ & $\zc$ & $+1.109271(55)$ & $+1.108275(52)$ & $+1.106822(31)$ & $+1.105005(28)$ & $+1.101902(24)$ \\
	$Z$ & $\zb$ & $+1.098592(37)$ & $+1.098350(42)$ & $+1.099529(24)$ & $+1.099629(20)$ & $+1.098754(18)$ \\
	\cmidrule(lr){1-7}
	$m$ & $0$   & $-0.68742(27)$  & $-0.68634(52)$  & $-0.68405(29)$  & $-0.68073(35)$  & $-0.67579(42)$  \\
	$m$ & $\zc$ & $-0.64955(20)$  & $-0.65657(27)$  & $-0.65970(21)$  & $-0.65998(24)$  & $-0.65982(29)$  \\
	$m$ & $\zb$ & $-0.523453(90)$ & $-0.55583(15)$  & $-0.57727(11)$  & $-0.59044(11)$  & $-0.60633(12)$  \\
	\bottomrule
\end{tabular}

%% file: tables/tab_kappas_phys.tex
\begin{tabular}[]{cclllll} \toprule 
$\zh$  & $\Delta$ & $\kappa_4(\zh, 12)$ 
                        & $\kappa_4(\zh, 16)$ 
                                           & $\kappa_4(\zh, 20)$ 
                                                              & $\kappa_4(\zh, 24)$ 
                                                                                & $\kappa_4(\zh, 32)$ \\ \midrule
$\zc$ & $\Del[c]$ & $0.1334348(15)$ & $0.1337542(15)$ & $0.1338369(13)$ & $0.1338351(10)$ & $0.1337351(11)$ \\
$\zb$ & $\Del[b]$ & $0.1229103(83)$ & $0.1264194(78)$ & $0.1282309(68)$ & $0.1293235(52)$ & $0.1304986(53)$ \\
\cmidrule(lr){1-7}
$\zc$ & $0$       & $0.1334308(15)$ & $0.1337524(16)$ & $0.1338361(13)$ & $0.1338346(10)$ & $0.1337349(11)$ \\
$\zb$ & $0$       & $0.121436(13)\ $ & $0.126049(13)\ $ & $0.1280872(72)$ & $0.1292528(53)$ & $0.1304734(55)$ \\
\bottomrule \end{tabular} 

%% file: conclusion.tex
\section{Conclusions and outlook}\label{s:conclusion}

We have carried out the renormalisation of massless $\nf=3$ QCD along three
lines of constant physics, defined by the gradient flow coupling with
Schr\"odinger functional boundary conditions. The specific choices are listed
in table~\ref{tab:summary}. Renormalisation factors of the flavour non-singlet
bilinears have been computed: axial-vector, pseudo-scalar, vector, tensor. Also
several improvement coefficients $b_X,\,c_X$ have been determined. In
table~\ref{tab:summary} we provide the pointers to tables and figures with
detailed results. 

Since our primary goal was to perform the renormalisations for the ALPHA
collabo\-ra\-tion B-physics project~\cite{Conigli:2023trw}, which is based on
step-scaling~\cite{Luscher:1991wu}, the lattice spacings are mostly rather
small. They are too small for direct large-volume computations, cf.
figure~\ref{fig:LCPs}. One might, however, devise step-scaling approaches
similar to~\cite{Conigli:2023trw} for applications other than B-physics.

A quick look at the ``precision'' column of the summary table reveals very high
to extremely high precision for quantities determined by Ward identities
(WIs). This can be understood theoretically. Variances are given by squaring
the path integral observable. For those quantities $Y$ determined by a WI,
written in the form $\langle W(Y) \rangle = 0$, the WI can be used in the path
integral representation of the variance. This removes the largest part of the
variance, leaving the remaining terms due to $\rmO(a^n)$ violations of the WI as well
as contact terms. We provide more details in appendix~\ref{app:variance}. 

Since chiral symmetry breaking by the Wilson term is restored by the chiral
WIs, one may also say that close to the continuum limit chiral symmetry
breaking plays no major role after renormalisation and $\rmO(a)$ improvement. Of
course, the determination of the various coefficients has to be carried out.
Here we have not determined the coefficients of $\rmO(a)$ terms that involve
the sea quark masses since these are not needed in the step-scaling functions
of~\cite{Conigli:2023trw}. Some of them can be determined through relations of
PCAC masses at finite and small sea quark masses. A general principle for
determining these coefficients is to impose the absence of linear terms in the
quark mass in a finite periodic volume~\cite{DallaBrida:2023fpl} or at short
distances~\cite{Korcyl:2016ugy}.

We have computed all necessary renormalisations with a precision that will be
subdominant in the physics observables of the B-physics
project~\cite{Conigli:2023trw,Conigli:2023rod,Kuberski:2026bpx} and expect that
selected results will be useful beyond this application. We have further
determined renormalisation and improvement coefficients in a partially massive
valence scheme in section~\ref{sec:massive}.
In addition to previous
implementations~\cite{deDivitiis:2019xla,Fritzsch:2010aw,Heitger:2003ue}, we
determined a full $\rmO(a)$ improved LCP for heavy valence quarks at the scale
of the heavy quark ($\Delta_{\rm h}=\Delta(\zh)$), which effectively includes
some relevant massive higher-order cutoff effects.
This scheme is tailored for B-physics, and we have demonstrated that it
does reduce higher-order discretisation errors in this context. It has contributed to the good
scaling observed in ref.~\cite{Kuberski:2026bpx}.

Finally, in our simulations we have varied the lattice spacing by a significant
factor, keeping physics scales constant. This allowed us to verify the expected
critical slowing down $\sim a^{-2}$ of the HMC algorithm, cf.
figures~\ref{fig:tauint} and~\ref{fig:tauint-L2} in appendix~\ref{a:ensembles}.

\begin{table}
    \centering	
    \small
    \begin{tabular}{@{\extracolsep{0.0cm}}cccccccc}\toprule
    LCP    & $L\,[{\rm fm}]$ & $\gbar^2$                       & results                & precision [\textperthousand] & table                  & figure                \\\midrule
    LCP-0  & 0.25            & $\gbsq_\mathrm{GF}(L_0)=3.949$  & $\za,\zp(L_0)$         & 0.03,~0.8                    & \ref{tab:ens_L0}       & \ref{fig:ZA_comb}     \\
           &                 &                                 & $Z,\,\ba-\bp,\,\bm$    & 0.02,\,2,\,0.02              & \ref{tab:RXall}        & \ref{fig:n_fit}       \\
           &                 &                                 & $Z_\mathrm{T}(L_0)$    & 1                            & \ref{tab:ZV}           &                       \\[0.5ex]
    LCP-0s & 0.5             & $\gbsq_\mathrm{GF}(L_0)=3.949$  & $\Sigma(\gbar^2(L_0)),\,\Sigma_\mathrm{P}(\gbar^2(L_0))$ 
                                                                                        &                              & \ref{tab:ens_2L0},\,\ref{tab:runs-stdobs}
                                                                                                                                                & \ref{fig:gfstep},\,\ref{fig:ZPstep}   \\
           &                 &                                 & $\cA, \,\zv,\,\bV$     &  5,\,0.02,\,0.4              & \ref{tab:ZV}           & \ref{fig:cA}          \\[1ex]
    LCP-1  & 0.5             & $\gbsq_\mathrm{GF}(L_1)=5.867$  & $\cA, \,\zp(L_1)$      & 5,\,1                        & \ref{tab:runs-stdobs}  & \ref{fig:cA}          \\[0.5ex]
    LCP-1s & 1.0             & $\gbsq_\mathrm{GF}(L_1)=5.867$  & $\Sigma(\gbar^2(L_1)),\,\Sigma_\mathrm{P}(\gbar^2(L_1))$ 
                                                                                        &                              & \ref{tab:ens_H2s},\,\ref{tab:runs-stdobs}
                                                                                                                                                & \ref{fig:gfstep},~\ref{fig:ZPstep}    \\
    \bottomrule
    \end{tabular}
    \caption{Summary of our main results. The computed $Z$-factors and
             improvement coefficients are specified in the ``results'' column
             and their numerical values are listed in the cited tables.
             Illustrations are found in the quotes figures.  We also quote rough
             numbers for the precision in some cases. It is very high for
             quantities determined by Ward identities. An explanation is
             provided in appendix \ref{app:variance}.
            }\label{tab:summary}
\end{table}

%% file: a_tabresfit.tex
\section{Tables of further results and fit formulae}\label{a:tables} 

\subsection{Tree-level normalisations}  

We list the normalisation factors ensuring
$\gbsq_\mathrm{GF}=g_0^2+\rmO(g_0^4)$ as well as $\ZP=1+\rmO(g_0^2)$  and
$Z_{\rm T}=1+\rmO(g_0^2)$ in tables \ref{tab:norm_g2GF}, \ref{tab:ZP_norm} 
and~\ref{tab:ZT_norm}.
\begin{table}[p!]
    \setlength{\abovecaptionskip}{2pt}
    \setlength{\belowcaptionskip}{10pt}
    \centering\small
    \renewcommand{\arraystretch}{1.25}
    \caption{Normalisation constant $\mathcal{N}(a/L)$ of the gradient flow 
             coupling $\gbsq_{\rm GF}(L)$, defined from the Zeuthen flow for
             the tree-level improved Lüscher-Weisz gauge action and observable
             at vanishing boundary fields.
            }
    \input{./tables/norm_g_GF.tex}
    \label{tab:norm_g2GF}
\end{table}
\begin{table}[p!]
	\setlength{\abovecaptionskip}{2pt}
	\setlength{\belowcaptionskip}{10pt}
	\centering\small
	\renewcommand{\arraystretch}{1.25}
    \caption{Summary of normalisation constant $c$ defined by $Z_\mathrm{P} =
             1$ at tree-level  with $T=L$, vanishing boundary fields $C=C'=0$
             and $\theta_k = \theta = 0.5$.
	        }
	\input{./tables/ctable_zp.tex}

	\label{tab:ZP_norm}
\end{table}
\begin{table}[p!]
	\setlength{\abovecaptionskip}{2pt}
	\setlength{\belowcaptionskip}{10pt}
	\centering\small
	\renewcommand{\arraystretch}{1.25}
    \caption{Summary of normalisation constant $c'$ defined by $Z_\mathrm{T} =
             1$ at tree-level  with $T=L$, vanishing boundary fields $C=C'=0$
             and $\theta_k = \theta = 0.5$.
	        }
	\input{./tables/ctable_zt.tex}
	\label{tab:ZT_norm}
\end{table}
\begin{table}[p!]
	\setlength{\abovecaptionskip}{2pt}
	\setlength{\belowcaptionskip}{10pt}
	\centering\small
	\renewcommand{\arraystretch}{1.25}
    \caption{Summary of the tree-level values for $\left[r^{(0)}(\theta) -
             r^{(0)}(\theta')\right]$ as used in eq.~(\ref{eq:cAdef}) for
             $\theta = 0.5$ and $\theta' = 0$ with $T=L$ and vanishing boundary
             fields $C=C'=0$.
	        }
	\input{./tables/cAtable_tree.tex}
	\label{tab:cA_tl}
\end{table}
%

\subsection{Ensembles of various LCPs}\label{sec:lcps}

The tuning parameters for $L_0$ lattices have been given in
table~\ref{tab:ens_L0} and results in doubled lattices in
table~\ref{tab:ens_2L0} in the main text.  We list our results for the tuning
to LCP-1 ($L_1$ lattices) and LCP-2 ($L_2$ lattices) in
tables~\ref{tab:ens_HQET1} and~\ref{tab:ens_HQET2}. Observables for
step-scaling lattices ($2L_1$), are listed in table~\ref{tab:ens_H2s}.

\begin{table}[p!]
	\setlength{\abovecaptionskip}{10pt}
	\setlength{\belowcaptionskip}{0pt}
	\centering\small
	\renewcommand{\arraystretch}{1.25}
    \caption{Ensembles tuned to {\bf LCP-1},
             $\gbsq_\mathrm{GF}(L)=5.867,\,m=0$, i.e., $L\approx L_1=2L_0$.
             All ensembles feature $T=L$. $N_{\rm cfg}$ is the total number of
             configurations, separated by $\tau_{\rm ms}$ MD units, on which the
             coupling and Schrödinger functional correlation functions are
             evaluated to compute the sea current quark mass $am=am_{12}$. The
             trajectory length is $\tau=2$\,MDU. $N_0$ denotes the number of
             configurations with vanishing topological charge.
	        }
	\input{./tables/enstab_HQET1.tex}
	\label{tab:ens_HQET1}
\end{table}
\begin{table}[p!]
	\setlength{\abovecaptionskip}{10pt}
	\setlength{\belowcaptionskip}{0pt}
	\centering\small
	\renewcommand{\arraystretch}{1.25}
    \caption{Ensembles tuned to {\bf LCP-1s}. Quantities as in
             table~\ref{tab:ens_HQET1}.
	        }
	\input{./tables/enstab_H2s.tex}
	\label{tab:ens_H2s}
\end{table}
\begin{table}[p!]
	\setlength{\abovecaptionskip}{10pt}
	\setlength{\belowcaptionskip}{0pt}
	\centering\small
	\renewcommand{\arraystretch}{1.25}
    \caption{Ensembles tuned to {\bf LCP-2},
             $\gbsq_\mathrm{GF}(L)=11.27,\,m=0$, i.e., $L\approx L_2=4L_0$.
             Quantities as in table~\ref{tab:ens_HQET1}.
	        }
	\input{./tables/enstab_HQET2.tex}
	\label{tab:ens_HQET2}
\end{table}

\subsection{Projection to \texorpdfstring{$Q=0$}{Q=0}}\label{sec:Q}

The topological charge is defined as in \cite{DallaBrida:2016kgh},
\begin{eqnarray}
   Q = \dfrac{1}{32 \pi^2}\,a^4\sum_x  \epsilon_{\mu\nu\rho\sigma} \mathrm{Tr}\left[G_{\mu\nu}^a(x,t)G_{\rho\sigma}^a(x,t)\right]^{\rm cl}_{\sqrt{8t}=0.3\,L}	\,,
\end{eqnarray}
with the clover definition, ``cl'', for the lattice field tensor at flow time
$\sqrt{8t}=cL$ with $c=0.3$, using the Zeuthen flow.  The projector is taken as
\begin{align}
     \hat{\delta} (Q) = \begin{cases}
	1 & |Q| <0.5\\
	0 & \text{otherwise}
    \end{cases}
    \,, \label{eq:projector_to_Qzero}
\end{align}
such that projected expectation values are
\begin{eqnarray}\label{eq:projection_to_Qzero}
	\langle {\cal O} \rangle_{Q=0} = {{\langle {\cal O} \, \hat{\delta} (Q)\rangle}\over{\langle  \hat{\delta} (Q)\rangle}} \,. 
\end{eqnarray} 
In \cite{Luscher:2011bx} it has been shown that polynomials of flowed gauge
fields do not require renormalisation (beyond the one of coupling and quark
mass).  One may worry that \eqref{eq:projection_to_Qzero} is not just a
polynomial in the fields, but it has further been demonstrated numerically that
configurations away from integer $Q$ are suppressed with a high power of $a$ in
the lattice path integral~\cite{Luscher:2010we}. Therefore
\eqref{eq:projection_to_Qzero} preserves the renormalisation properties of
$\langle {\cal O} \rangle$.

%% file: tables/norm_g_GF.tex
\begin{tabular}[t]{cccccccc}\toprule
$L/a$ & $\mathcal{N}(a/L)$ & $L/a$ & $\mathcal{N}(a/L)$ & $L/a$ & $\mathcal{N}(a/L)$    \\\cmidrule(lr){1-2}\cmidrule(lr){3-4}\cmidrule(lr){5-6}\cmidrule(lr){7-8}
  4    & 0.0123411705 & 14 & 0.0086111544 & 32 & 0.0085655417 \\
  6    & 0.0101626915 & 16 & 0.0085917584 & 40 & 0.0085644807 \\
  8    & 0.0090316148 & 20 & 0.0085753596 & 48 & 0.0085640980 \\
 10    & 0.0087449664 & 24 & 0.0085693878 & 64 & 0.0085638539 \\
 12    & 0.0086509179 & 28 & 0.0085668037 &    &              \\
 \bottomrule
\end{tabular}

%% file: tables/ctable_zp.tex
\begin{tabular}[t]{clclclcl}\toprule
 $L/a$ & $c$           & $L/a$ & $c$           & $L/a$ & $c$          & $L/a$ & $c$        \\\cmidrule(lr){1-2}\cmidrule(lr){3-4}\cmidrule(lr){5-6}\cmidrule(lr){7-8}
 $6 $  & $0.98969662$ &  $24$ & $0.99934941$ & $42$  & $0.99978746$ & $60$ &$0.99989584$\\
 $8 $  & $0.99417664$ &  $26$ & $0.99944559$  & $44$  & $0.99980634$ & $62$ &$0.99990246$\\
 $10$  & $0.99626479$  &  $28$ & $0.99952193$  & $46$  & $0.99982281$ & $64$ &$0.99990846$\\
 $12$  & $0.99740297$ &  $30$ & $0.99958352$  & $48$  & $0.99983727$ & $68$ &$0.99991891$\\
 $14$  & $0.99809059$  &  $32$ & $0.99963393$ & $50$  & $0.99985002$ & $72$ &$0.99992767$\\    
 $16$  & $0.99853742$ &  $34$ & $0.99967572$  & $52$  & $0.99986134$ & $96$ &$0.99995931$\\    
 $18$  & $0.99884400$  &  $36$ & $0.99971074$  & $54$  & $0.99987142$ &      &            \\ 
 $20$  & $0.99906343$  &  $38$ & $0.99974038$  & $56$  & $0.99988044$ &      &            \\ 
 $22$  & $0.99922584$  &  $40$ & $0.99976568$  & $58$  & $0.99988854$ &      &            \\ 
 \bottomrule
\end{tabular}

%% file: tables/ctable_zt.tex
\begin{tabular}[t]{clclclcl}\toprule
$L/a$ & $c'$ & $L/a$ & $c'$ & $L/a$ & $c'$ & $L/a$ & $c'$\\
\cmidrule(lr){1-2}\cmidrule(lr){3-4}\cmidrule(lr){5-6}\cmidrule(lr){7-8}
$6$ & $-1.39063503$ & $24$ & $-1.40194109$ & $42$ & $-1.40245685$ & $60$ & $-1.40258458$\\
$8$ & $-1.39587394$ & $26$ & $-1.40205428$ & $44$ & $-1.40247909$ & $62$ & $-1.40259237$\\
$10$ & $-1.39831978$ & $28$ & $-1.40214413$ & $46$ & $-1.40249850$ & $64$ & $-1.40259945$\\
$12$ & $-1.39965453$ & $30$ & $-1.40221664$ & $48$ & $-1.40251553$ & $68$ & $-1.40261177$\\
$14$ & $-1.40046162$ & $32$ & $-1.40227600$ & $50$ & $-1.40253057$ & $72$ & $-1.40262210$\\
$16$ & $-1.40098643$ & $34$ & $-1.40232521$ & $52$ & $-1.40254390$ & $96$ & $-1.40265942$\\
$18$ & $-1.40134673$ & $36$ & $-1.40236646$ & $54$ & $-1.40255578$ &  & \\
$20$ & $-1.40160470$ & $38$ & $-1.40240137$ & $56$ & $-1.40256642$ &  & \\
$22$ & $-1.40179572$ & $40$ & $-1.40243118$ & $58$ & $-1.40257597$ &  & \\
\bottomrule
\end{tabular}

%% file: tables/cAtable_tree.tex
\begin{tabular}[t]{clclcl}
\toprule
$L/a$ & $r^{(0)}(\theta) - r^{(0)}(\theta')$        & $L/a$ & $r^{(0)}(\theta) - r^{(0)}(\theta')$       & $L/a$ & $r^{(0)}(\theta) - r^{(0)}(\theta')$         \\
\cmidrule(lr){1-2}\cmidrule(lr){3-4}\cmidrule(lr){5-6}
$8$  & $2.28728\times10^{-5}$ & $20$ & $5.85873\times10^{-7}$ & $40$ & $3.66183\times10^{-8}$ \\
$12$ & $4.51980\times10^{-6}$ & $24$ & $2.82548\times10^{-7}$ & $48$ & $1.76567\times10^{-8}$ \\
$16$ & $1.43028\times10^{-6}$ & $32$ & $8.94031\times10^{-8}$ & $64$ & $5.58769\times10^{-9}$ \\
\bottomrule
\end{tabular}

%% file: tables/enstab_HQET1.tex
\begin{tabular}{@{\extracolsep{0.0cm}}ccllcccc}\toprule
$L/a$ & $\beta$  & $\kappa$ & $\bar{g}^2_\mathrm{GF}(L)$ 
                                                & $L\,m$     & $N_{\rm cfg}$ & $N_{0}$ 
                                                                       & $\dfrac{\tau_{\rm ms}}{\rm MDU}$    \\ \midrule
     8  & $3.6537$ & $0.1370722$ & $5.8648(65)$ & $+0.00195(79)$ & $15000$ & $14977$ &  4 \\
    12  & $3.8349$ & $0.1369654$ & $5.8697(87)$ & $-0.00118(48)$ & $15004$ & $14992$ &  4 \\
    16  & $4.0018$ & $0.1366803$ & $5.865(11) $ & $+0.00036(32)$ & $18000$ & $17999$ &  4 \\
    20  & $4.1394$ & $0.1363857$ & $5.857(11) $ & $+0.00148(20)$ & $22020$ & $22018$ &  4 \\
    24  & $4.2530$ & $0.1361224$ & $5.877(14) $ & $+0.00216(19)$ & $20002$ & $20002$ &  4 \\ 
\bottomrule
\end{tabular}

%% file: tables/enstab_H2s.tex
\begin{tabular}{@{\extracolsep{0.0cm}}ccllcccc}\toprule
$L/a$ & $\beta$  & $\kappa$ & $\bar{g}^2_\mathrm{GF}(L)$ 
                                                & $(L/2)\,m$     & $N_{\rm cfg}$ & $N_{0}$ 
                                                                       & $\dfrac{\tau_{\rm ms}}{\rm MDU}$    \\ \midrule
    16  & $3.6537$ & $0.1370722$ & $9.377(25) $ & $+0.02232(30) $ & $15000$ & $12213$ &  8  \\
    24  & $3.8349$ & $0.1369654$ & $10.485(36)$ & $+0.00957(15) $ & $15000$ & $12423$ &  16 \\
    32  & $4.0018$ & $0.1366803$ & $10.888(43)$ & $+0.007405(73)$ & $72000$ & $61178$ &  4  \\
    40  & $4.1394$ & $0.1363857$ & $10.971(59)$ & $+0.006821(73)$ & $27120$ & $27081$ &  4  \\
    48  & $4.2530$ & $0.1361224$ & $11.279(76)$ & $+0.007050(47)$ & $56400$ & $56342$ &  4  \\ 
\bottomrule
\end{tabular}

%% file: tables/enstab_HQET2.tex
\begin{tabular}{@{\extracolsep{0.0cm}}cclllrrc}\toprule
$L/a$ & $\beta$  & $\kappa$ & $\bar{g}^2_\mathrm{GF}(L)$
                                                & $L\,m$     & $N_{\rm cfg}$ & $N_{0}$
                                                                       & $\dfrac{\tau_{\rm ms}}{\rm MD}$    \\ \midrule
  12  & $3.4014$ & $0.1368240$ & $11.288(57)$ & $-0.0002(24) $ &  $5004$ & $3319$ &  4 \\
  14  & $3.4798$ & $0.1370410$ & $11.244(41)$ & $-0.0013(14) $ &  $19200$ & $12122$ &  4 \\
  16  & $3.5522$ & $0.1371379$ & $11.299(55)$ & $+0.0004(13) $  &  $6000$ & $4014$ &  4 \\
  20  & $3.6900$ & $0.1371452$ & $11.227(49)$ & $+0.00025(66)$  & $20600$ & $15392$ &  4 \\
  24  & $3.8013$ & $0.1370387$ & $11.248(55)$ & $-0.00042(38)$  & $20000$ & $13941$ &  4 \\
  28  & $3.8989$ & $0.1368897$ & $11.195(68)$ & $-0.00044(35)$  & $11200$ & $11061$ & 4 \\
  32  & $3.9764$ & $0.1367450$ & $11.269(98)$ & $+0.00005(26)$  & $15059$ & $13659$ & 4 \\
\bottomrule
\end{tabular}

%% file: a_compZA.tex
\section{Comparison of different definitions of \texorpdfstring{$\ZA$}{ZA}}\label{sec:ambiguitiesZA}

\begin{figure}
    \centering
    \includegraphics[width=14cm]{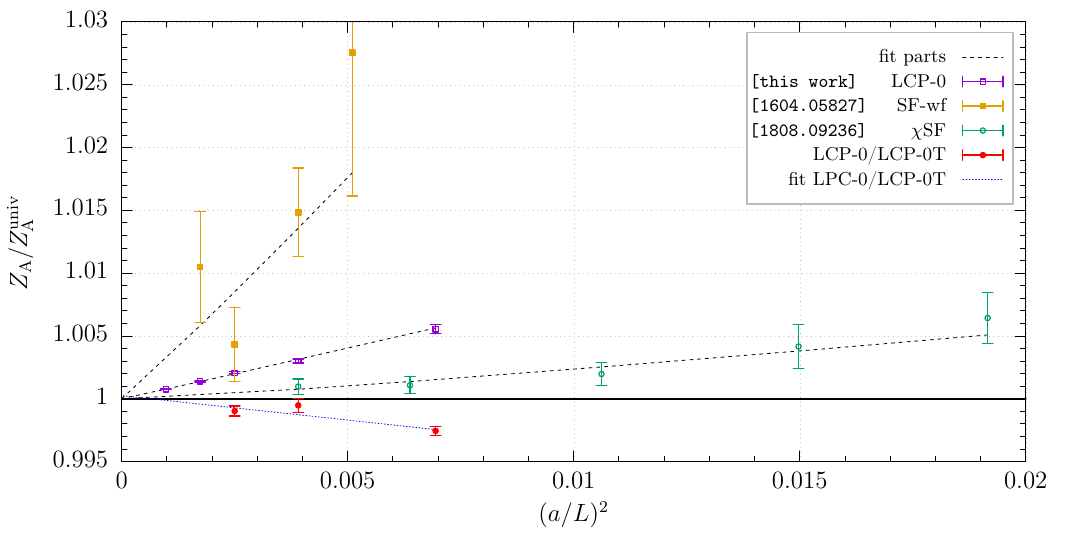}
    \caption{$\ZA$ normalised to the universal fit plotted against the
             $(a/L)^2$. For the various definitions, see the main text. Note
             that each definition has a rather different $L$ in physical units.
             The points labeled ''LCP-0/LCP-0T'' show $\za(T=L)/\za(T=2L)$.
            }\label{fig:ZA_diff}
\end{figure}

Figure~\ref{fig:ZA_diff} shows $\ZA$ for the  various definitions normalised to
the universal fit of eq.~\eqref{e:zauniv}. In this way one can see the quality
of the fit and the fact that indeed the data is compatible with the universal
function to within a percent for LCP-0 and the $\chi$SF (chirally rotated SF)
results, and also the SF results of ref.~\cite{Bulava:2016ktf} differ only by
less than 3\%.
%
\begin{table}[t!]
	\setlength{\abovecaptionskip}{2pt}
	\setlength{\belowcaptionskip}{10pt}
	\centering\small
	\renewcommand{\arraystretch}{1.25}
    \caption{Ensembles {\bf LCP-0T} ($L=L_0$) created for an explicit check of
            the ambiguities in the determinations of $\za$. We list the
            associated results for the (fixed) gradient flow coupling, sea
            quark masses and renormalisation constant $\za$. The norms for
            calculating the gradient flow coupling change due to the change of
            the time extent from $T/L=1$ to $T/L=2$ and read
            $(0.0070065025,0.0069879795,0.0069828385)$ for $T/a=(24,32,40)$,
            respectively. Otherwise, the definition of $\gbsq_{\rm GF}$ is the
            same as detailed in eq.~\eqref{eq:g2GF} and
            table~\ref{tab:norm_g2GF}.
	        }
    \label{t:bXz_hqet_enstab_T2L}%
    \label{tab:ensL02T}%
	\input{./tables/enstab_T2L.tex}
\end{table}
%
Furthermore, we checked that our choice for $T/L$ with resulting smallish
distance of insertion points $x_0$ and $y_0$ from the boundary does not
introduce  significantly enhanced discretisation errors.  We generated three
ensembles of sizes $L_0 \in \left\{12, 16, 20\right\}$ with $T=2L_0$. These
ensembles, together with the measured sea quark masses and the values for the
renormalisation constant $Z_\mathrm{A}$, are summarised in
table~\ref{t:bXz_hqet_enstab_T2L}.
The lower part of figure~\ref{fig:ZA_diff} shows the ratio
$Z_\mathrm{A}(T=L)/Z_\mathrm{A}(T=2L)$ also labeled as LCP-0/LCP-0T, together
with a linear fit to the expected functional behaviour, $1+l_1 (a/L)^2$. It can
be seen that the ratio (lower curve) is sufficiently small and compatible with
a quadratic behaviour towards the continuum limit. We  conclude that the time
extent of the LCP-0 lattices is large enough.
%

%% file: tables/enstab_T2L.tex
\begin{tabular}{@{\extracolsep{0.0cm}}clllrlllll}
	\toprule
  $L/a$ & $T/a$& $\beta$& $\kappa$  & $N_{\rm r}$ 
                                           & $\dfrac{\tau_\mathrm{ms}}{\mathrm{MD}}$ 
                                                &  $N_{\rm cfg}$ 
                                                       & $\bar{g}^2_\mathrm{GF}$  
                                                                      & $L\,m$      & $Z_\mathrm{A}$ \\\cmidrule(lr){1-7}\cmidrule(lr){8-10}
      12  &  24  & 4.3030 & 0.1359947 &  5 &  8 & 5321 & $4.2024(76)$ & $0.00676(46)$ & $0.83393(30)$  \\
      16  &  32  & 4.4662 & 0.1355985 &  2 & 10 & 3461 & $4.2251(99)$ & $0.00473(42)$ & $0.83926(46)$  \\
      20  &  40  & 4.6017 & 0.1352848 & 10 & 10 & 2000 & $4.214(15)$  & $0.00499(43)$ & $0.84571(33)$  \\
    \bottomrule
\end{tabular}

%% file: vectorcurrent.tex
\section{Renormalisation of vector and tensor currents}\label{a:zvt}

In order to compute form factors for semi-leptonic B-meson decays within the
ALPHA collaboration strategy for
B-physics~\cite{Bernardoni:2013xba,Conigli:2023trw,Conigli:2023rod}, the
renormalisation of the corresponding matrix elements is to be performed in a
small volume, i.e., for observables defined on the LCP-0s ensembles.  This
primarily concerns the vector current and, in the case of radiative decays,
also the tensor current~\cite{Meinel:2024pip}.  We compute the corresponding
renormalisation factors on the ensembles of LCP-0s and LCP-0, respectively.

As in previous work, see e.g.~\cite{Heitger:2020zaq,Fritzsch:2018zym}, we use
the vector Ward identity
\begin{align} \label{e:ZV_WI_amq}
	Z_{\rm V} \left(1 + b_{\rm V} a m_{\rm q} + \bar{b}_{\rm V} \mathrm{Tr}[aM_{\rm q}] \right)
	= {{F_1}\over{F_{\rm V}}(T/2)} \,,
\end{align}
with $F_1$ as defined in eq.~(\ref{e:f1}) and
\begin{align}
        F_{\rm V}(x_0) = {{a^3}\over{2}}\sum_{\mathbf{x}}\rmi \epsilon^{abc} 
	\langle \mathcal{O}'^{a} V_0^{b}(x_0, \mathbf{x}) \mathcal{O}^{c} \rangle \,.
\end{align}
First, we determine the improvement coefficient
$b_{\rm V}$ from the slope
\begin{align}
	b_{\rm V}
	=
	\left.
	{{\partial}\over{\partial (a m_{\rm q})}}
	\log\!\left[{{F_1}\over{F_{\rm V}(T/2)}}\right]
	\right|_{m_{\rm q}=0} \, .
\end{align}
For this purpose, we use measurements at three different values of
$a m_{\rm q}$.
In addition to the unitary point $m_{45}=m_{12}$, we perform two further
measurements with reduced statistics at slightly smaller valence quark masses
$m_{45}$; see table~\ref{tab:massshifts}.
The valence quark masses are chosen such that the point of vanishing PCAC mass
lies within the covered range.
Linear fits of the form
$\log(F_1/F_{\rm V}) = p + b_{\rm V}\, (2\kappa_4)^{-1}$ have a good quality of fit and directly determine $b_{\rm V}$.

Second, using the renormalisation factor $Z$ (as given in table~\ref{tab:RXall} for
massless valence quarks), we rewrite eq.~(\ref{e:ZV_WI_amq}) to contain PCAC
masses,
\begin{align} \label{e:ZV_WI_amij}
	Z_{\rm V} 
	= {{F_1}\over{F_{\rm V}}(T/2)} \left(1 + {{b_{\rm V}}\over{Z}} a m_{45} + {{\bar{b}_{\rm V}}\over{Z}} a(m_{12} + m_{23} + m_{31}) \right)^{-1}\,.
\end{align}

\begin{table}[t!]
	\setlength{\abovecaptionskip}{10pt}
	\setlength{\belowcaptionskip}{0pt}
	\renewcommand{\arraystretch}{1.25}
	\centering\small
    \caption{Valence PCAC quark masses $(L/2)\cdot m_{45}$ at two nearby values
             of the hopping parameter $\kappa_4 = \kappa_5$ for each ensemble on LCP-0s.
             Combined with the unitary quark mass shown in
             table~\ref{tab:ens_2L0}, these measurements allow for an
             interpolation to vanishing valence quark mass. The additional data
             are obtained with reduced statistics.
	        }\label{tab:massshifts}
	\input{./tables/tab_m45_shifted}
\end{table}

We evaluate $Z_{\rm V}$ in this form
at the unitary point using
the previously determined value of $b_{\rm V}$, the factor $Z$ from
table~\ref{tab:RXall}, and $\bar{b}_{\rm V}$ from~\cite{Fritzsch:2018zym}. The
results for $Z_{\rm V}$ and $b_{\rm V}$ are summarised in table~\ref{tab:ZV},
where we observe a very high level of statistical precision.

The non-singlet tensor current is scale-dependent. Its renormalisation group
running has been computed in~\cite{Chimirri:2023ovl}. Once the renormalisation
factor is determined at a given scale, it can be obtained at other scales through the
running factor. In analogy to the pseudo-scalar density, we compute $Z_{\rm
T}$ at the scale $\mu = 1/L_0$, i.e., on the ensembles of LCP-0.

Following~\cite{Chimirri:2023ovl}, we define
\begin{align}
	Z_{\rm T}^{\tt f} &= c' \, {{\sqrt{F_1}}\over{k_{\rm T}^{\rm I}(T/2)}} \,, & 
	Z_{\rm T}^{\tt k} &= c' \, {{\sqrt{K_1}}\over{k_{\rm T}^{\rm I}(T/2)}} \,,
\end{align}
where the constant $c'$ is chosen such that $Z_{\rm T}=1$ at tree level. We
collect $c'$ for different lattice sizes in table~\ref{tab:ZT_norm}.

In analogy with $F_1$ in eq.~\eqref{e:f1}, we define the boundary-to-boundary
correlation function in the vector channel
\begin{align}
	K_1\equiv-{{1}\over{24}}\,\sum_{i=1}^2\langle{\mathcal{Q}}_k'^{\,3i} {\mathcal{Q}}_k^{\,i3}\rangle \,,
\end{align}
where a sum over the repeated spatial index $k=1,2,3$ is implied, and the
boundary operators are given by
\begin{align}
    {\mathcal{Q}}^{ji}_k  &= \dfrac{a^6}{L^3} \sum_{\vecu,\vecv}\zetabar_{j}({\vecu}) \gamma_k \zeta_i({\vecv}) \,, &
	{\mathcal{Q}}'^{ji}_k &= \dfrac{a^6}{L^3} \sum_{\vecu,\vecv}\zetabar'_j({\vecu}) \gamma_k \zeta'_i({\vecv}) \,. 
\end{align}
The $\rmO(a)$ improved boundary-to-bulk correlation function $k^{12}_{\rm
T,\mathrm{I}}$ of the tensor current
\begin{align}
    T^{12}_{\mu\nu}(x)&= \rmi\psibar_1(x) \sigma_{\mu\nu}\psi_2(x) \;, &
    \text{with}~\sigma_{\mu\nu}   &=\frac{\rmi}{2}[\gamma_\nu,\gamma_\mu] \;,
\end{align}
is defined as
\begin{align}
        k^{12}_{\rm T,\mathrm{I}}(x_0) = k^{12}_{\rm T}(x_0) + a c_{\rm T}\tilde \partial_0 k^{12}_{\rm V}(x_0) \,,
\end{align}
where 
\begin{align}
        k_{\rm T}^{12}(x_0) &= -{{1}\over{6}} a^3 \sum_{\mathbf{x}} \langle T_{0k}^{12}(x) \mathcal{Q}_k^{21} \rangle \,,& 
        k_{\rm V}^{12}(x_0) &= -{{1}\over{6}} a^3 \sum_{\mathbf{x}}	\langle V_k^{12}(x) \mathcal{Q}_k^{21} \rangle \,.
\end{align}
The improvement coefficient $c_{\rm T}$ is taken from the interpolating formula
given in eq.~(3.22) of~\cite{Chimirri:2023ovl}.

Our results for the tensor renormalisation factor are listed in
table~\ref{tab:ZV}. At this stage, we do not include a systematic uncertainty
$\sim a/L$ associated with the incomplete knowledge of the boundary improvement
coefficient $\tilde{c}_t$, as proposed in eq.~(4.17)
of~\cite{Chimirri:2023ovl}. However, on our coarsest ensemble on LCP-0, we
find that the estimated dependence on $\tilde{c}_t$ is consistent with the
expectation of~\cite{Chimirri:2023ovl}, in the sense that we reproduce the
predicted derivative with respect to $\tilde{c}_t$ at the level of about
$20\%$. It is then recommended to take this uncertainty estimate into account
when performing continuum extrapolations of matrix elements of $T_{\mu\nu}$
with $\rmO(a^2)$ corrections.
\begin{table}[t!]
	\setlength{\abovecaptionskip}{10pt}
	\setlength{\belowcaptionskip}{0pt}
	\renewcommand{\arraystretch}{1.25}
	\centering\small
    \caption{Table of renormalisation constants (and mass-improvement
             coefficient $b_{\rm V}$) for the vector and tensor currents as
             obtained in this work on the ensembles of the LCP-0 ($Z_{\rm
             T}^{\tt f}$, $Z_{\rm T}^{\tt k}$) and LCP-0s ($Z_{\rm V}$, 
             $b_{\rm V}$) trajectories.
            }\label{tab:ZV}
	\input{./tables/tab_renorm_VT}
\end{table}

%% file: tables/tab_m45_shifted.tex
\begin{tabular}{@{}cc cc cc@{}}
	\toprule
	& & \multicolumn{2}{c}{set 1} & \multicolumn{2}{c}{set 2} \\
	\cmidrule(lr){3-4}\cmidrule(lr){5-6}
	$L/a$ & $\beta$ & $\kappa_4$ & $(L/2)\cdot m_{45}$ & $\kappa_4$ & $(L/2)\cdot m_{45}$ \\
	\midrule
	24 & 4.3030 & 0.1360087 & $-0.00083(24)$ & 0.1360227 & $-0.00589(24)$ \\
	32 & 4.4662 & 0.1356118 & $0.00173(22)$ & 0.1356165 & $-0.00054(22)$ \\
	40 & 4.6017 & 0.1352886 & $0.00160(20)$ & 0.1352925 & $-0.00076(20)$ \\
	48 & 4.7165 & 0.1350207 & $0.00083(20)$ & 0.1350232 & $-0.00099(20)$ \\
	64 & 4.9000 & 0.1346036 & $0.00306(17)$ & 0.1346081 & $-0.00132(17)$ \\
	\bottomrule
\end{tabular}

%% file: tables/tab_renorm_VT.tex
	\begin{tabular}{@{\extracolsep{0.0cm}}cccccc}\toprule
            $L/a$ &   $\beta$ & $Z_{\rm V}$   & $b_{\rm V}$   & $Z_{\rm T}^{\tt f}(L)$   & $Z_{\rm T}^{\tt k}(L)$   \\
		\midrule
    12 &    4.3030 & 0.807546(16)  & 1.31293(45)   & 1.06039(98)     & 1.00330(75)     \\
	16 &    4.4662 & 0.817176(10)  & 1.28975(52)   & 1.0857(16)      & 1.0270(12)      \\
	20 &    4.6017 & 0.8245733(84) & 1.27362(48)   & 1.1057(18)      & 1.0450(13)      \\
	24 &    4.7165 & 0.8304271(99) & 1.26036(55)   & 1.1190(21)      & 1.0582(16)      \\
	32 &    4.9000 & 0.8390158(64) & 1.24155(50)   & 1.1379(29)      & 1.0751(22)      \\
		\bottomrule
	\end{tabular}

%% file: variance.tex
\section{On the variance of PCAC quark masses}\label{app:variance}

Here we present a qualitative discussion of the variance of the quark masses,
as an example of a quantity computed from Ward identities. Ignoring details
such as $Q=0$ projection, improvement, (and initially) the sum over different
time slices, we rewrite the condition for the PCAC mass, $m$, as
\begin{eqnarray}
 F(m)\stackrel{!}{=} 0,\quad \text{with}\quad  F(m) = \langle W^{12}(x_0)\rangle\,, 
\end{eqnarray}
where
\begin{eqnarray}
     W^{12}(x_0) = a^3\sum_{\bf  x} w^{12}(x) \,\mathcal{O}^{21}\,, \quad w^{12}(x)=
	 {\partial}_0  A^{12}_0(x) - 2 m\,P^{12}(x)\,,
\end{eqnarray}
and note that the external operators ${\mathcal{O}}^{ji}={a^6\over L^3}
\sum_{\bf x,y}\zetabar_{j}({\bf x}) \gamma_5 \zeta_i({\bf y})$ have vanishing
mass dimension, $[{\mathcal{O}}^{ji}]=0$, and create only space-momentum zero
states. The variance $\Sigma^2(m)$ of $m$ is related to that of $F$ by
\begin{eqnarray}
	\Sigma^2(m) \cdot (F'(m))^2 = \Sigma^2(F(m)) ,
\end{eqnarray}
where $F'(m)$ is the sensitivity of the WI to $m$.  We are from now on
interested only in the qualitative behaviour at small $a$, i.e. the dependence
on powers of $a$. In other words, we ignore multiplicative $a$-independent or
only logarithmically varying terms and denote this by the $\sim$ symbol. Since
$F'(m) = \rmO(a^0)$ we  have
\begin{eqnarray}
	\Sigma^2(m) \sim \Sigma^2(F(m))\,.
\end{eqnarray}

The variance related to fluctuations of only the gauge field is given by 
\begin{eqnarray}
  \Sigma^2(F) &=& \langle [W^{12}(x_0)]_\psi  [W^{12}(x_0)]_\psi\rangle_U - \langle [W^{12}]_\psi\rangle_U^2 \nonumber\\
  &=& \langle [W^{12}(x_0)]_\psi  [W^{12}(x_0)]_\psi\rangle_U \,,
	\label{e:Grassmann}
\end{eqnarray}
where $[\bullet]_\psi$ is the (exact) Grassmann ``average'' over the fermion
fields and $\langle.\rangle_U$ the gauge average. Introducing a quenched copy
of the fermions with indices 4,5, we can avoid splitting fermion and gauge
average, and write
\begin{eqnarray}
	\Sigma^2(F) &=& \langle W^{12}(x_0)\,  W^{45}(x_0)\rangle \,
	\nonumber\\
	&=& a^6\sum_{\bf  x,x'}
	\langle 
	w^{12}(x_0,{\bf x}) \,w^{45}(x_0,{\bf x'})\,\mathcal{O}^{21}\,\mathcal{O}^{54} 
	\rangle \,. 
	\label{e:sig1}
\end{eqnarray}
Close to the continuum limit, we may use the PCAC operator relation
$w^{12}(x)\approx 0$ at all points except $\bf x = \bf x'$ and get
\begin{eqnarray}
	\Sigma^2(F) 	&\sim& a^6\sum_{\bf  x}
	\langle 
        a^{d-8}\ \mathcal{O}^{d}(x_0,{\bf x}) \,\mathcal{O}^{21}\,\mathcal{O}^{54} 
	\rangle \,, 
	\label{e:sig2}
\end{eqnarray}
where $\mathcal{O}^{d}(x_0,{\bf x})$ is a dimension $d$ Lorentz-scalar operator
which can arise from mixing with $w^{12}(x_0,{\bf x}) \,w^{45}(x_0,{\bf x})$.
Since higher-dimensional operators are suppressed by powers of $a$, only the
lowest $d$ is relevant. Due to the (non-singlet) flavour structure
$\psibar_1\psi_2\,\psibar_4\psi_5$ of $\mathcal{O}^{d}$ we have $d=6$. Using
translation invariance 
\begin{eqnarray}
	 a^3\sum_{\bf  x}
	\langle 
	\mathcal{O}^{d}(x_0,{\bf x}) \,\mathcal{O}^{21}\,\mathcal{O}^{54} 
	\rangle 
	&=& L^3
	\langle 
	\mathcal{O}^{d}(x_0,{\bf x}) \,\mathcal{O}^{21}\,\mathcal{O}^{54} 
	\rangle \,,
	\label{e:sig3}
\end{eqnarray}
we obtain
\begin{eqnarray}
	\Sigma^2(LF) ~ \sim ~ a^{d-5} ~ \sim ~ {a \over L} \,.
\end{eqnarray}
The last equation is just due to dimensional reasoning and therefore valid up to
some dimensionless function $f(g_0)$ for small $m$. Clearly, to get a
non-vanishing contribution in perturbation theory, we need two gluon lines,
connecting the 12 and 54 parts. This provides a factor $g_0^4$ and we find
\begin{eqnarray}\label{e:sig4a}
	\Sigma^2(LF) \sim g_0^4 {a \over L} \,.
\end{eqnarray}
Finally, let us take into account the average over $|I_{T}|$ values of $x_0$ in
eq.~\eqref{e:x0average}. The contact term in the variance only appears for
equal-time contributions, which also means that the correlation between
estimates $m(x_0)$ at $x_0$ with those at $y_0\ne x_0$ vanishes to the
accuracy considered here. Thus we get another suppression
\begin{eqnarray}\label{e:sig4}
    \Sigma^2(Lm_{\eqref{e:x0average}}) \sim g_0^4 {a \over L} \dfrac{1}{|I_{T}|}\sim g_0^4 {a^2 \over LT}\,.
\end{eqnarray}
We note that in several steps, e.g. in \eqref{e:Grassmann} or going from
\eqref{e:sig1} to \eqref{e:sig2}, we have used the PCAC relation dropping the
$\rmO(a^2)$ terms. Thus \eqref{e:sig4a} and \eqref{e:sig4} are a priori
expected to hold only close to the continuum.

\begin{figure}
	\centering
	\includegraphics[width=.8\textwidth]{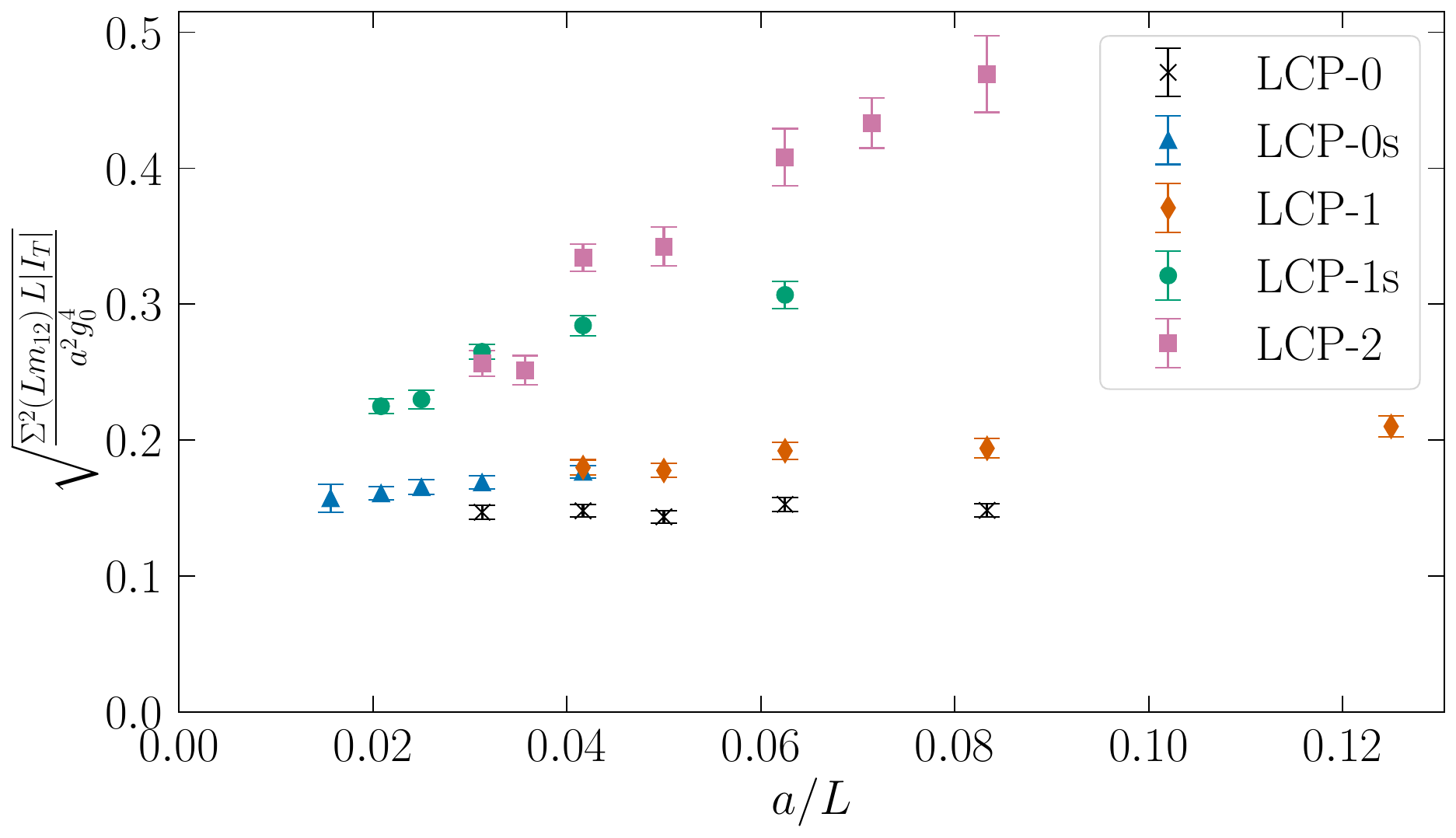}
    \caption{Standard deviation of the sea quark mass, as evaluated on the
             gauge ensembles of the five LCPs used in this work, normalised by
             the predicted leading-order behaviour from eq.~\eqref{e:sig4}.
	        }\label{fig:variance}
\end{figure}

We verify the expected scaling behaviour numerically by studying the variance
of the sea-quark PCAC masses defined via eq.~\eqref{e:x0average} on our
ensembles. In figure~\ref{fig:variance}, we show the square root of the
variance, rescaled by the leading prediction of eq.~\eqref{e:sig4}. At small
values of the coupling, the rescaled data point to a finite, non-zero value
for the LCP-0, LCP-0s and LCP-1 ensembles.  At larger coupling, we observe
deviations which we attribute to cutoff effects.%
\footnote{The scaling with $g_0^2$ at fixed $a/L$ is also seen in the data of
        \cite{Fritzsch:2018zym}. Assuming that autocorrelation effects are
        mild, this again verifies the above prediction.
}
The above argument can also be made for other quantities determined from WIs.
However, it is important that the fermion ``average''  \eqref{e:Grassmann} does
not lead to flavour singlet correlators (quark line disconnected), since
otherwise  the contact term may be highly divergent.
Finally, we note that for small $a/L$ very high precision is indeed observed
numerically in our observables determined from WIs, such as $\za,\zv,Z$,
while, e.g., $\zp,Z_\mathrm{T}$ are more than an order of magnitude less
precise.

%% file: app_sim.tex
\section{Gauge ensembles \label{a:ensembles}}

\subsection{Action}  

As already mentioned, we use the same discretisation as in
ref.~\cite{DallaBrida:2016kgh}. The discretised action is specified in section
II.B and eq.~(3.6) of \cite{DallaBrida:2016kgh}.

\subsection{Algorithmic details, critical slowing down, topology freezing}

In this section we summarise some details of the main simulations which led
to the results presented in this paper. These details are organised by the
different lines of constant physics, cf.\ tables \ref{tab:ens_L0},
\ref{tab:ens_2L0}, and the tables in section~\ref{sec:lcps}.  We use the same
setup that was chosen in ref.~\cite{DallaBrida:2016kgh} with the RHMC
algorithm for the third, massless quark. We refer to this paper also for the
exact details of the action, including the SF boundary improvement terms. 

Monte Carlo chains of lengths $N_{\rm cfg}\cdot\tau_{\rm ms} = \rmO(100\,000\,\mathrm{MDU})$ have been generated with the algorithmic parameters of
table~\ref{tab:runs-algo}. In table~\ref{tab:runs-stdobs} we list results for
$\gbsq_{\rm GF}(L)$, $\ZP$ and $F_1$ together with the integrated
autocorrelation time $\tau_{\rm int}$ in units of MDU. 
%
\begin{figure}[t!]
	\centering
	\includegraphics[width=.48\textwidth]{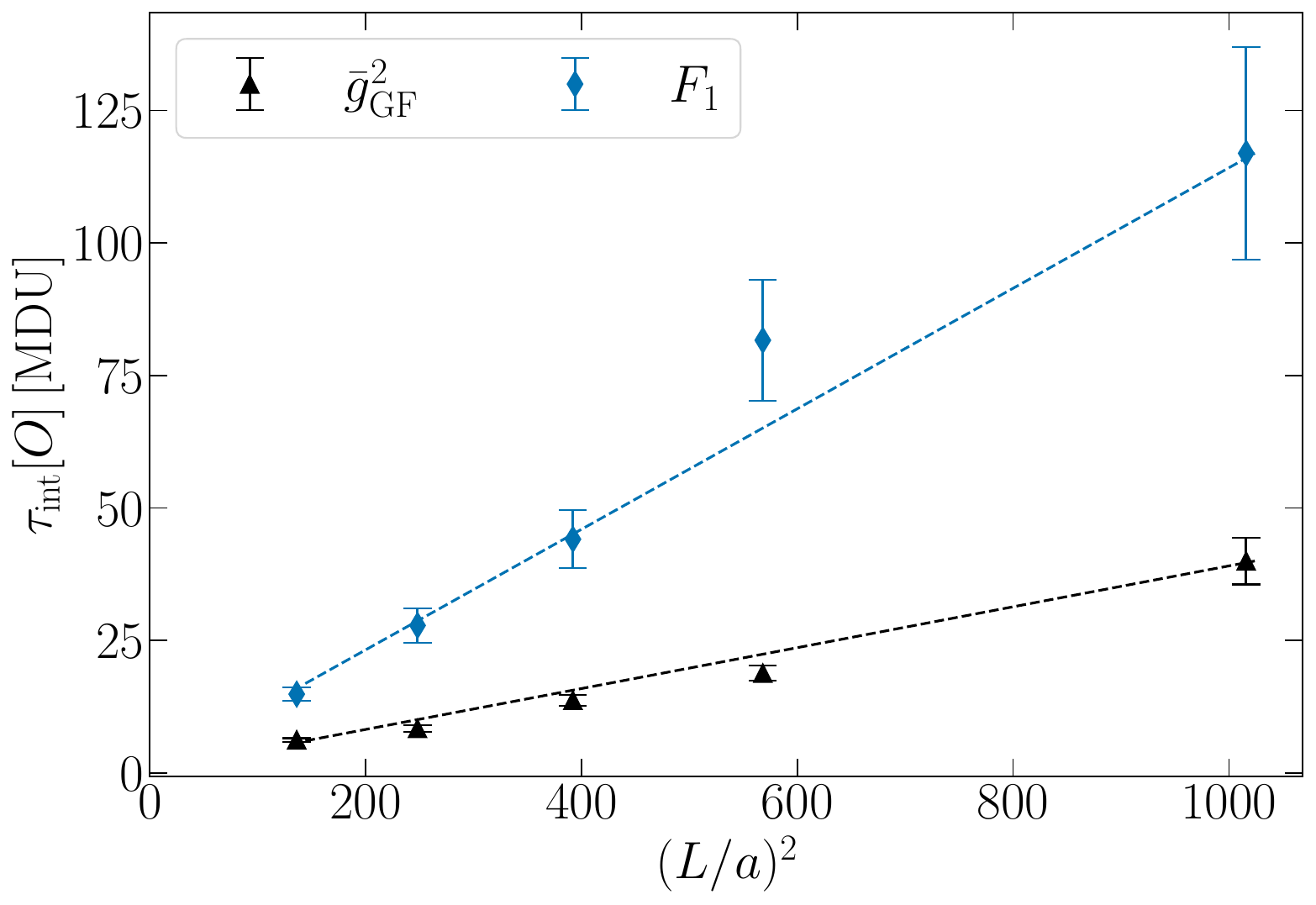}\hfill%
	\includegraphics[width=.48\textwidth]{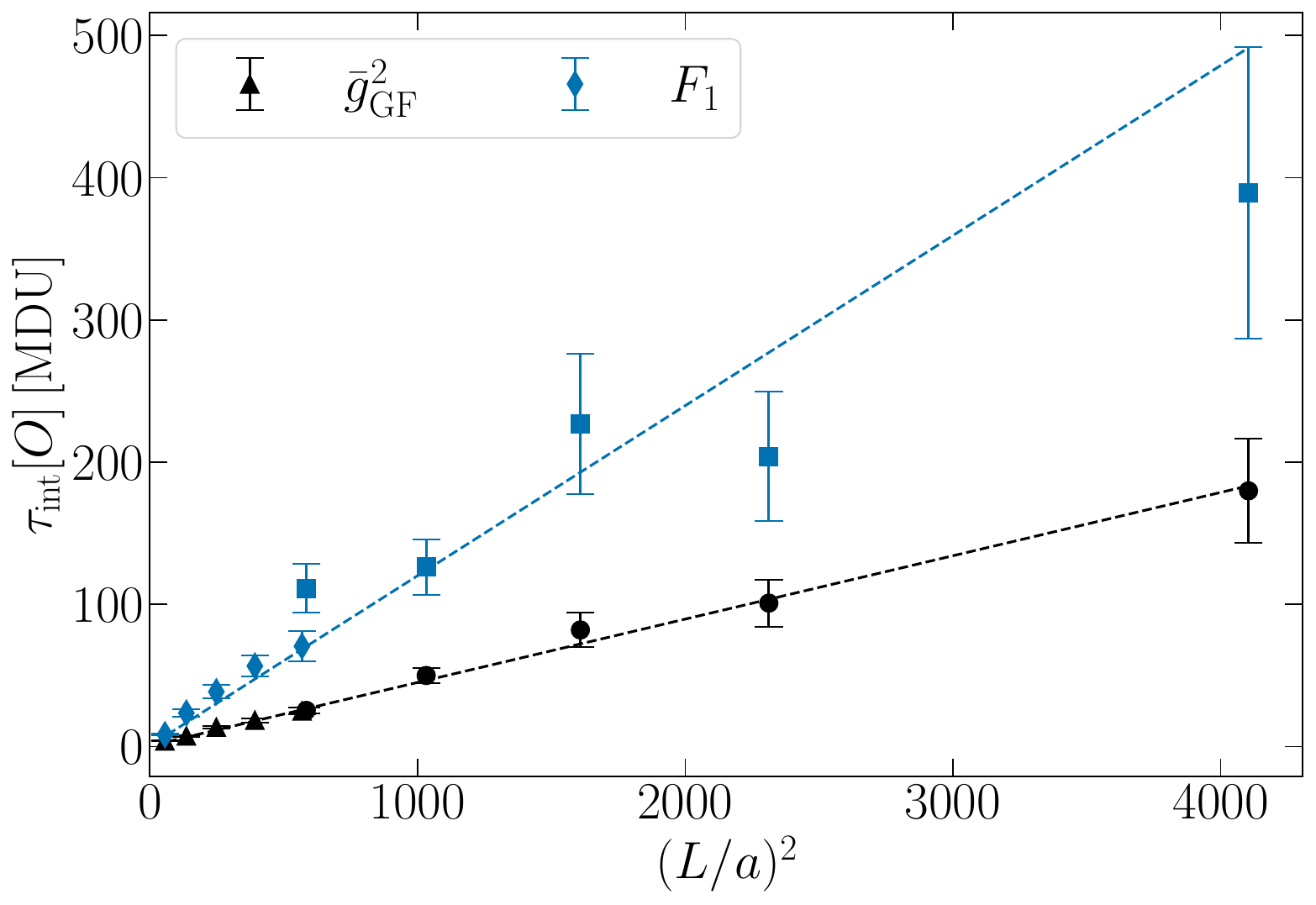}%
    \caption{Integrated autocorrelation times $\tau_{\rm int}$ of the gradient
             flow coupling $\gbsq_{\rm GF}$ and the boundary-to-boundary
             correlation function $F_1$. We show lattices with a spatial extent
             of about 0.25\,fm (\textit{left}) and 0.5\,fm (\textit{right})
             which are both in the frozen-charge situation.	Lines, based on
             a fit to the three largest ensembles each, are shown to guide the
             eye. On the right, LCP-1 (LCP-0s) simulations are located at
             smaller (larger) $L/a$ and distinguished by different symbols.
	        }\label{fig:tauint}
\end{figure}
\begin{figure}[t!]
	\centering
	\includegraphics[width=.48\textwidth]{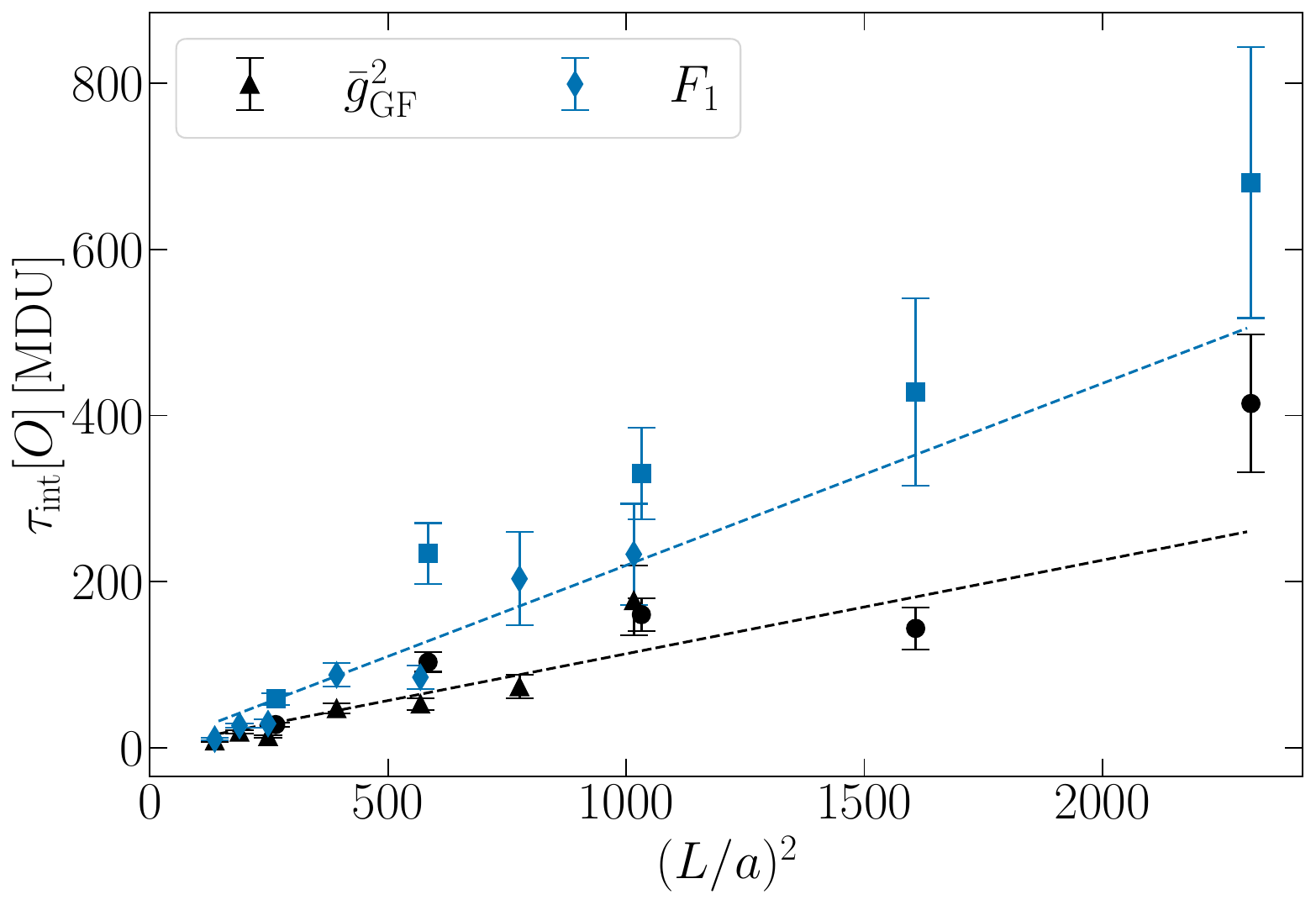}%
    \caption{As fig.~\ref{fig:tauint} but with a spatial extent of about 1\,fm
             where $1-N_0/N_\mathrm{cfg}$ reaches up to about $1/3$. LCP-2
             (LCP-1s) simulations are located at smaller (larger) $L/a$ and
             distinguished by different symbols.
            }\label{fig:tauint-L2}
\end{figure}
%
In all cases under consideration here, the integrated autocorrelation time
increases significantly, when approaching the continuum limit. We observe
critical slowing down. 

A special source for autocorrelation is the freezing of the topological charge
at small lattice spacing~\cite{DelDebbio:2002xa,Schaefer:2010hu}. Charges that
differ from zero are exponentially suppressed in small
volumes~\cite{Luscher:1981zf}. This helps to some extent, but for the $L_2$
lattices we face a more challenging situation. Here, non-zero charges do
occur with a probability of a few tens of percent, as clearly seen on the
coarser lattices where rather frequent tunnellings are observed. But the
sampling of the topological charge strongly depends on the lattice spacing.
The charge is frozen to zero on the very fine lattices of the ensembles tuned
to LCP-0 and LCP-0s in tables~\ref{tab:ens_L0} and~\ref{tab:ens_2L0}.  
Because we define our observables in the zero charge sector, cf.
eq.~\eqref{eq:projection_to_Qzero}, the frozen situation is not the difficult
one. Most problematic are the ensembles in the intermediate region, where the
charge changes once in a while and the simulation may be stuck in sectors of
non-zero charge for a number of trajectories. This can have a negative effect
on the autocorrelation times. 

It is then reassuring that all in all the observables defined by
eq.~\eqref{eq:projection_to_Qzero} show rather standard critical slowing down,
as illustrated for the ones with the largest autocorrelations in
figures~\ref{fig:tauint},~\ref{fig:tauint-L2}. Note that the freezing of the
charge happens around $a\approx 0.05\,\fm$
\cite{DelDebbio:2002xa,Schaefer:2010hu}. For $L\leq 0.5\,\fm$, shown in
fig.~\ref{fig:tauint}, all lattices have spacings below $0.05\,\fm$ and complete
agreement with $\tau_\mathrm{int}\propto a^{-2}$ is observed. On the other
hand, for $L\approx1\,\fm$ the transition to freezing is at $L/a \approx 20$.
There we observe some disagreement between estimates for
$\tau_\mathrm{int}[F_1]$ on LCP-2 and $\tau_\mathrm{int}[F_1]$ on LCP-1s
despite these LCPs being very close. We interpret this as a sign of
insufficient statistics to properly determine the uncertainty of
$\tau_\mathrm{int}$ in the difficult transition region. We emphasise again that
in the figures we showed the observables with the strongest autocorrelations.

\begin{table}[t!]
	\setlength{\abovecaptionskip}{5pt}
	\setlength{\belowcaptionskip}{2pt}
	\centering\small
	\renewcommand{\arraystretch}{1.25}
    \caption{Algorithmic parameters of our major simulations. The column
             labelled ``RHMC'' lists the lower and upper limits of the
             approximation range and the number of poles in the Zolotarev
             approximation. The next column gives information concerning the
             relative variation of the corresponding reweighting factor.  We
             also list the average plaquette, the acceptance rate and
             expectation value of the exponentiated Hamiltonian deficit at
             the end of the trajectories.
	        }\label{tab:runs-algo}
	\input{./tables/runs_algo.tex}

\end{table}

\begin{table}[t!]
	\setlength{\abovecaptionskip}{5pt}
	\setlength{\belowcaptionskip}{2pt}
	\centering\small
	\renewcommand{\arraystretch}{1.25}
    \caption{Some standard observables and their integrated autocorrelation
             times in units of MDU.
            }\label{tab:runs-stdobs}
	\input{./tables/runs_stdobs.tex}

\end{table}

%% file: tables/runs_algo.tex
\begin{tabular}[]{cclllll} \toprule
        LCP   & $L/a$& RHMC               &  $\dfrac{\sigma_{W}}{\langle W\rangle}$  
                                       & $p_\mathrm{avg}$ & $P_\mathrm{acc}$ & $\langle e^{-\Delta H}\rangle$ \\ \midrule
 LCP-0  & $12$ & $[0.0900,6.0]$:7   & $1.71 \cdot 10^{-5}$ & 1.991048(21) & 96\% & 1.00036(47) \\
        & $16$ & $[0.0700,6.1]$:8   & $1.38 \cdot 10^{-5}$ & 2.015524(12) & 99\% & 0.99990(14) \\
        & $20$ & $[0.0600,5.9]$:8   & $2.35 \cdot 10^{-5}$ & 2.0377882(68) & 90\% & 1.0033(12) \\
        & $24$ & $[0.0500,5.9]$:8   & $5.61 \cdot 10^{-5}$ & 2.0571320(37) & 95\% & 0.99991(51) \\
        & $32$ & $[0.0300,5.9]$:11  & $0.29 \cdot 10^{-5}$ & 2.0879275(23) & 92\% & 0.99869(89) \\
        \cmidrule(lr){1-7}
 LCP-0s & $24$ & $[0.0415,6.0]$:8   & $6.56 \cdot 10^{-5}$ & 1.9588131(45) & 93\% & 0.99981(72) \\
        & $32$ & $[0.0300,5.9]$:9   & $4.69 \cdot 10^{-5}$ & 1.9923208(22) & 95\% & 0.99924(46) \\
        & $40$ & $[0.0280,5.9]$:10  & $2.28 \cdot 10^{-5}$ & 2.0198645(16) & 93\% & 1.00005(84) \\
        & $48$ & $[0.0220,5.9]$:10  & $12.28 \cdot 10^{-5}$ & 2.0426170(11) & 95\% & 1.00068(57) \\
        & $64$ & $[0.0130,5.9]$:14  & $0.30 \cdot 10^{-5}$ & 2.07750159(56) & 95\% & 1.00049(59) \\
                                               \midrule
 LCP-1  & $8$  & $[0.0300,6.4]$:7  & $6.00 \cdot 10^{-5}$ & 1.831287(73) & 97\% & 0.99981(44) \\
        & $12$ & $[0.0200,6.4]$:8  & $6.46 \cdot 10^{-5}$ & 1.855370(30) & 94\% & 0.99954(88) \\
        & $16$ & $[0.0200,6.2]$:9  & $2.49 \cdot 10^{-5}$ & 1.890274(14) & 96\% & 1.00013(47) \\
        & $20$ & $[0.0450,6.0]$:8  & $2.87 \cdot 10^{-5}$ & 1.9205457(74) & 95\% & 1.00015(53) \\
        & $24$ & $[0.0220,5.9]$:8  & $17.10 \cdot 10^{-5}$ & 1.9454846(56) & 93\% & 0.99889(92) \\
        \cmidrule(lr){1-7}
 LCP-1s & $16$ & $[0.0120,7.5]$:10 & $2.53 \cdot 10^{-5}$ & 1.771909(14) & 96\% & 0.99982(48) \\
        & $24$ & $[0.0100,6.5]$:10 & $6.25 \cdot 10^{-5}$ & 1.8180137(39) & 91\% & 0.99782(84) \\
        & $32$ & $[0.0075,6.5]$:11 & $2.44 \cdot 10^{-5}$ & 1.8637227(19) & 93\% & 0.9995(15) \\
        & $40$ & $[0.0160,6.1]$:10 & $4.95 \cdot 10^{-5}$ & 1.9001788(18) & 95\% & 1.00092(64) \\
        & $48$ & $[0.0100,6.0]$:11 & $3.12 \cdot 10^{-5}$ & 1.92904823(86) & 90\% & 0.9965(13) \\
        \midrule
 LCP-2  & $12$ & $[0.0150,7.8]$:9  & $6.66 \cdot 10^{-5}$ & 1.692340(68) & 99\% & 1.00004(15) \\
        & $14$ & $[0.0120,7.5]$:9  & $6.17 \cdot 10^{-5}$ & 1.712718(32) & 98\% & 1.00036(27) \\
        & $16$ & $[0.0120,7.5]$:10 & $4.09 \cdot 10^{-5}$ & 1.733092(26) & 92\% & 0.9975(17) \\
        & $20$ & $[0.0150,7.0]$:10 & $2.11 \cdot 10^{-5}$ & 1.774058(11) & 96\% & 0.99873(72) \\
        & $24$ & $[0.0100,6.5]$:10 & $5.38 \cdot 10^{-5}$ & 1.8064704(53) & 88\% & 0.9951(20) \\
	& $28$ & $[0.0085,7.5]$:11 & $4.71 \cdot 10^{-5}$ & 1.8344256(57) & 99\% & 0.99997(24) \\
        & $32$ & $[0.0075,7.5]$:12 & $5.30 \cdot 10^{-5}$ & 1.8557454(42) & 92\% & 0.9968(16) \\
\bottomrule 
\end{tabular}

%% file: tables/runs_stdobs.tex
\begin{tabular}[]{ccllllllll} \toprule 
 LCP    & $L/a$& $\gsqgf(L)$ & $\taui[\gsqgf]$  &  $\ZP(g_0^2,a/L)$ & $\taui[\ZP]$  & $F_1$ & $\taui[F_1]$  \\ \midrule
 LCP-0  & $12$ &   3.9461(43) &     6.22(33) &  0.57835(32) &     7.46(45) &   0.6069(18) &    14.9(1.2) \\
        & $16$ &   3.9475(58) &     8.38(60) &  0.56972(44) &    11.62(97) &   0.5702(25) &    27.8(3.3)  \\
        & $20$ &   3.9493(63) &    13.7(1.0) &  0.56503(50) &    23.0(2.2) &   0.5438(25) &    44.1(5.5) \\
        & $24$ &   3.9492(62) &    18.8(1.4) &  0.56002(47) &    29.5(2.7) &   0.5262(29) &       82(11) \\
        & $32$ &   3.9490(97) &    40.0(4.4) &  0.55388(69) &    52.2(6.5) &   0.4972(34) &      117(20)\\\cmidrule(lr){1-8}
LCP-0s & $24$ &    5.592(11) &    25.4(2.2) &  0.48080(49) &    25.0(2.1) &   0.4210(34) &      111(17) \\
        & $32$ &    5.689(14) &    49.9(5.3) &  0.47313(78) &    67.5(8.1) &   0.3916(30) &      126(20) \\
        & $40$ &    5.755(22) &       82(12) &  0.46677(92) &       76(11) &   0.3742(45) &      227(49) \\
        & $48$ &    5.770(25) &      101(17) &   0.4631(12) &      103(18) &   0.3568(44) &      204(46) \\
        & $64$ &    5.844(32) &      180(37) &   0.4553(19) &      250(57) &   0.3319(55) &     389(102) \\
        \midrule
 LCP-1  & $8$  &   5.8648(65) &     3.84(19) &  0.49392(31) &     4.00(20) &   0.5940(20) &     8.26(58) \\
        & $12$ &   5.8697(87) &     7.39(49) &  0.47644(43) &     8.31(60) &   0.5142(28) &    23.5(2.6) \\
        & $16$ &    5.865(11) &    13.5(1.1) &  0.47349(48) &    13.2(1.1) &   0.4649(30) &    38.5(4.8) \\
        & $20$ &    5.857(11) &    18.4(1.5) &  0.47051(61) &    25.4(2.4) &   0.4348(29) &    56.6(7.5) \\
        & $24$ &    5.877(14) &    24.9(2.5) &  0.46808(81) &    40.1(4.8) &   0.4142(34) &       71(11) \\\cmidrule(lr){1-8}
 LCP-1s & $16$ &    9.377(25) &    27.9(2.4) &   0.3471(11) &    46.7(5.1) &   0.3147(32) &    58.6(7.0) \\
        & $24$ &   10.485(36) &      103(12) &   0.3168(18) &      214(33) &   0.2425(39) &      234(37) \\
        & $32$ &   10.888(43) &      160(20) &   0.3043(20) &      304(49) &   0.2038(36) &      330(55) \\
        & $40$ &   10.971(59) &      144(25) &   0.3013(33) &     397(102) &   0.1886(56) &     428(113) \\
        & $48$ &   11.279(76) &      415(83) &   0.2909(31) &     592(135) &   0.1667(46) &     680(163) \\\midrule
 LCP-2  & $12$ &   11.288(57) &     8.21(94) &   0.2897(16) &    10.7(1.3) &   0.4408(66) &    10.4(1.3) \\
 		& $14$ &   11.244(41) &    19.0(1.7) &   0.2903(13) &    25.5(2.6) &   0.3721(44) &    26.7(2.8) \\
        & $16$ &   11.299(55) &    13.6(1.7) &   0.2870(22) &    23.5(3.7) &   0.3177(70) &    29.0(5.0) \\
        & $20$ &   11.227(49) &    47.1(6.0) &   0.2937(19) &       76(12) &   0.2694(49) &       88(14) \\ 
        & $24$ &   11.248(55) &    52.7(7.1) &   0.2953(23) &       89(15) &   0.2391(46) &       85(14) \\
    	& $28$ &   11.195(68) &       74(14) &   0.2961(33) &      149(37) &   0.2143(72) &      204(56) \\
        & $32$ &   11.269(98) &      177(42) &   0.2971(39) &      252(68) &   0.2145(65) &      233(61) \\

\bottomrule 
\end{tabular}

%% file: Bss-renorm.bbl
\providecommand{\href}[2]{#2}\begingroup\raggedright\endgroup